\documentclass{llncs}
\usepackage{latexsym}
\usepackage{amsfonts}
\usepackage{amssymb}
\usepackage{epsfig}
\usepackage{times}

\begin{document}

\title{Pushdown Timed Automata:\\
a Binary Reachability Characterization\\
and Safety Verification\thanks{A short
 version \cite{D01} 
 of this paper
appears in the {\em Proceedings of the 13th
International Conference on Computer-aided
Verification} (CAV'01),
Lecture Notes in Computer Science 2102, pp. 506-517,
Springer.
}
}
\author{ 
Zhe Dang}
\institute{School of Electrical Engineering and Computer Science\\
Washington State University\\
Pullman, WA 99164, USA\\
{\tt zdang@eecs.wsu.edu}
}
\maketitle
\begin{abstract}
We consider
pushdown timed automata (PTAs) that are timed automata
(with dense clocks) augmented with a pushdown stack.
A configuration of a  PTA
includes a control state,  dense clock values
and a stack word.
By using
the pattern technique,
we give a decidable
characterization of
the binary reachability (i.e., the set of
all pairs of configurations such that
one can reach the other) of a
PTA.
Since a timed automaton  can be treated as
a PTA
  without the pushdown stack,
we can show that the binary reachability of a timed automaton
is definable in the additive theory of reals and integers.
The results can be used to verify a class
 of properties containing linear relations
over both dense variables and unbounded discrete variables.
The properties 
previously could not be
verified using the classic region technique
nor expressed by timed temporal logics  for timed automata
and
CTL$^*$ for pushdown systems. The results are also extended to
other generalizations of timed automata.
\end{abstract}

\newcommand{\N} {{\bf N}}
\newcommand{\Z} {{\bf Z}}
\newcommand{\R} {{\bf R}}
\newcommand{\Q} {{\bf Q}}
\newcommand{\D} {{{\bf D}^+}}
\newcommand{\F} {{{\bf D}}}

\newcommand{\deq} {{\sim}}
\newcommand{\rdeq} {{\approx}}
\def\relv #1{\widehat{#1}}

\def\up #1{{\lceil{#1}\rceil }}
\def\down #1{{\lfloor{#1}\rfloor }}
\newcommand{\srto}{\leadsto^{\cal A}_{\langle s,X_r\rangle}}
\newcommand{\srleads}{\leadsto^{\cal A}_{X_r}}
\def\B{{\bf B}}
\def\A{{\cal A}}
\def\C{{\cal C}}
\def\P{{\cal P}}
\def\RPTA{{\cal B}}

\def \v{{\vec v}}
\def \x{{\vec x}}
\def \u{{\vec u}}
\def \ND{{\bf (N, D)}-formula}
\def \FNPCM {{(\bf \F+NPCA)}}
\def\ring #1{{\stackrel{\circ}{#1}}}
\def\qed{\hfill\rule{1.5mm}{1.5ex}}

\section{
Introduction}\label{chapt0}
A timed automaton \cite{AD94} can be considered as
a finite automaton augmented with a number of dense (either real or rational)
clocks. Clocks
can be reset or progress at rate 1
depending 
upon the truth values of a number of 
clock constraints in the form of clock regions
(i.e., comparisons of a clock or the difference of two clocks 
against an integer constant). 
Due to their 
ability to model and analyze
 a wide range of real-time
systems, timed automata have been 
extensively studied in recent years (see \cite{A99,Y98} for recent surveys).
In particular,
by using the standard region technique,
it has been shown  that 
region reachability for  timed automata
is decidable \cite{AD94}.
This fundamental result and the technique
help researchers, both theoretically and practically,
in formulating various timed temporal logics
\cite{ACD93,AFH96,AH93,AH94,HNSY94,LLW95,RS97,W94}
and developing verification tools \cite{BDM98,HH95,LPY97}.

Region reachability is useful but has intrinsic limitations.
In many real-world applications \cite{CGK97}, we might also want to
know whether a timed automaton satisfies a non-region property,
e.g., 
$$x_1-2x_2+x_3'>x_1'+4x_2'-3x_3$$
 holds whenever
clock values $(x_1,x_2,x_3)$ can reach $(x_1',x_2',x_3')$.
Recently,
Comon and Jurski \cite{CJ99}
have shown that the binary reachability
of
a timed automaton is definable in the additive theory
of reals augmented with an integral predicate that
tells
whether a term is
an integer,
by flattening a timed automaton into
a real-valued counter machine without
nested cycles \cite{CJ98}.
The result immediately paves the way
 for
automatic verification of a class of
non-region properties 
that previously
 were not possible using the region technique.

On the other hand,
a strictly more
powerful system, called a {\em pushdown timed  automaton} (PTA),
 can be obtained
 by augmenting
a timed automaton with
a pushdown stack.
PTAs are particularly interesting because
they contain both dense clocks and
unbounded discrete structures. They 
can be
used to study, for instance,
a timed version
of pushdown processes \cite{BEM97,FWW97}
or real-time programs with procedure calls.
A configuration of a 
PTA is a tuple of
a control state, dense clock values, and a stack word.
The binary reachability 
of a PTA is the set of all pairs of
configurations such that
one can reach the other.
Comon and Jurski's result for timed automata
inspires us to
look for a similar result for
PTAs.
Is there 
a decidable
binary reachability characterization for 
PTAs such that a class of
non-region properties
can be verified ?
 The main result in this paper answers
this question positively.

There are several potential ways to approach the question.
The first straightforward
 approach
would be to
treat
a PTA as
a Cartesian product of
a timed automaton and a pushdown automaton.
In this way, the binary reachability of
a PTA can be formulated
by simply combining Comon and Jurski's result and 
the fact that pushdown automata accept context-free languages.
Obviously, this is wrong, since stack operations
depend on clock values and thus can not be simply separated. 
The second approach is
to closely
look
at the flattening technique of Comon and Jurski's
  to see
whether  the technique can be adapted
 by adding a pushdown stack.
However, the second approach has
an  inherent difficulty:
the flattening technique, as pointed out
in their paper,  destroys the structure of the original
timed automaton,  and thus, the sequences of stack operations can not
be maintained after flattening. 

Very recently, 
the question
has been answered positively, 
but only for integer-valued clocks
(i.e., for discrete PTAs).
It has been shown in \cite{DIBKS00}
that the binary reachability
of a discrete PTA
can be accepted by a
 nondeterministic pushdown  automaton
augmented with reversal-bounded counters (NPCA),
whose emptiness problem is known to be decidable \cite{I78}.
However, as far as dense clocks are concerned, the 
automata-based technique used in 
\cite{DIBKS00} does not apply. The reason
is that traditional automata theories
do not provide tools to deal with machines
containing both
real-valued
counters (for dense clocks)
  and unbounded discrete data structures.

In order to handle dense clocks,
we introduce a new technique, called 
the pattern technique, by separating a dense clock into an integral part and
a fractional part.
Consider a pair $(\v_0, \v_1)$ of two tuples of clock values.
We define (see Section \ref{pattern} for details)
an ordering, called the pattern of
$(\v_0, \v_1)$,
 on the fractional parts
of $\v_0$ and $\v_1$.
The definition guarantees that there are only
a finite number of
distinct patterns.
An equivalent relation
``$\rdeq$" is defined such that
$(\v_0, \v_1)\rdeq (\v_0',\v_1')$ iff
$\v_0$ and $\v_0'$ ($\v_1$ and $\v_1'$ will also)
have the same integral parts, and
both $(\v_0, \v_1)$ and $(\v_0',\v_1')$ have the same pattern.
The ``$\rdeq$" essentially defines an equivalent relation
with a countable number of   equivalent classes such that
the integral parts of $\v_0$ and $\v_1$ together
with the pattern of the fractional parts
of $\v_0$ and $\v_1$ determine
 the equivalent class of 
$(\v_0, \v_1)$.
A good property of ``$\rdeq$" is 
that 
it preserves the binary reachability:
$\v_0$ can reach $\v_1$ by a sequence of transitions
iff $\v_0'$ can reach $\v_1'$ by the (almost)
same sequence of transitions, whenever
$(\v_0, \v_1)\rdeq (\v_0',\v_1')$.
Therefore, the fractional parts can be abstracted away
from the dense clocks by using a pattern.
In this way, by preserving the (almost) same control structure,
a PTA
can be transformed into 
a discrete transition system (called a pattern graph)
containing discrete clocks (for the integral parts
of the dense clocks) and a finite variable over patterns.
By translating
a pattern back to a relation over the
fractional parts of the clocks,
the  decidable binary reachability
characterization of 
 the pattern graph  derives
the decidable characterization (namely, $\FNPCM$-definable)
 for the PTA, since
the relation is definable in the additive theory of reals.
With this characterization,
it can be shown that the particular
 class of safety properties that 
contain mixed linear relations
over both dense variables (e.g., clock values)
 and discrete variables (e.g., word counts)
can be automatically verified for PTAs.
For instance,
\begin{quote}
whenever configuration $\alpha$ can reach
configuration $\beta$, $\alpha_{x_1}+2\beta_{x_2}-\alpha_{x_2}
>\#_a(\alpha_{\bf w})-\#_b(\beta_{\bf w})$ holds.
\end{quote}
can be verified, where $\alpha_{x_1}$ is
 the dense value for
clock $x_1$ in $\alpha$, $\#_a(\alpha_{\bf w})$ is the number of 
symbols $a$ in the stack word
 of $\alpha$.
The results can be easily extended to PTAs
augmented with reversal-bounded counters.
In particular, 
 we can show that the
binary reachability of a timed automaton is 
definable in the first-order additive theory over reals
and integers
 with $\ge$ and $+$, i.e., 
$(\R, \N, +, \ge, 0)$.
Essentially, for
timed automata,
Comon and Jurski's characterization (the additive theory of reals
augmented with an integral predicate)
is equivalent to 
 ours (the additive theory of reals and integers).
The additive theory over reals
and integers is decidable, for instance, by
the Buchi-automata based decision
procedure presented in \cite{BRW98}.

Fractional orderings are an effective way to abstract
the fractional parts of dense clocks. The idea of 
using fractional orderings can be traced back to the pioneering
work of Alur and Dill in inventing the region technique
\cite{AD94}. 
Essentially,  the region technique
makes a finite partition 
of the clock space such that clock values in the 
same region give the same answer to each clock constraint
in the system (i.e., the automaton of interest).
Comon and Jurski \cite{CJ99} notice that
Alur and Dill's partition is too coarse in establishing 
the binary reachability of a timed automata.
They move one step further by bringing in the clock values
before a transition was made.
But Comon and Jurski's partition is still finite,
since their partition, though finer than
Alur and Dill's, is still based on 
answers  to all the  clock constraints (there are finitely many of them)
in the system.
In this paper, $\rdeq$ deduces
an {\em infinite} partition of
both the initial values $\v_0$ and the current values $\v_1$
 of the clocks. Essentially,
this partition is 
based on 
answers  to all clock constraints (not just the ones in the system).
That is,
$\rdeq$ is finer
than Comon and Jurski's partition as well as 
Alur and Dill's. 
This is why the flattening technique \cite{CJ99}
destroys the transition structure of a timed automaton
but the technique presented in this paper is able to
preserve the transition structure.
A class of Pushdown Timed Systems
was discussed in \cite{BER95}. However, that paper
focuses on region reachability instead of binary reachability.

This paper is organized as follows.
Section \ref{prel} reviews a number of definitions and,
in particular, defines a decidable formalism in which
 the binary reachability
of PTAs are expressed.
Section \ref{pattern} and Section \ref{CCP}
 give the definition of patterns and show 
the correctness of
using patterns as an abstraction for fractional clock
values.
Section \ref{PTA} and 
Section \ref{graph}
 define PTAs
and show that 
the pattern graph of a PTA
has a decidable binary reachability characterization.
Section \ref{verif} states the main results of the paper.
In Section \ref{concl},
we point out that the results in this paper
can be extended to many other infinite state machine models
augmented with dense clocks.

\section{Preliminaries}\label{prel}
A nondeterministic multicounter automaton is
a nondeterministic automaton with
a finite number of states, 
a one-way input tape, and
a finite number of integer counters.
Each counter can be incremented by 1,
decremented by 1, or stay unchanged.
Besides, a counter can be tested against 0.
It is well-known that counter machines with two counters
have an
undecidable halting problem, and obviously the undecidability holds
for machines augmented with a pushdown stack.
Thus, we have to restrict
the behaviors of the counters. One such restriction is to
limit  the number of  reversals a counter can make.
A counter is {\it $n$-reversal-bounded} if it changes mode between
nondecreasing and nonincreasing at most $n$ times.
For instance, the
following sequence of  counter values:
$$0,0,1,1,2,2,3,3,4,4,3,2,1, 1, 1, 1,\cdots$$
demonstrates only one counter reversal.
A counter is {\it reversal-bounded} if it is $n$-reversal-bounded
for some fixed number $n$ independent of  computations.
A {\em reversal-bounded nondeterministic multicounter automaton (NCA)}
is a nondeterministic multicounter automaton in which
each counter is reversal-bounded.
A {\em reversal-bounded
nondeterministic
pushdown multicounter automaton (NPCA)} 
 is 
an NCA augmented with
a pushdown stack.
In addition to  counter operations,
an NPCA can pop the
top symbol from the stack
or push a word onto the top of the stack.
It is known that the emptiness problem (i.e., whether
a machine accepts some words?) for NPCAs (and
hence NCAs) is decidable.
\begin{lemma}\label{emptiness}
The emptiness problem for 
reversal-bounded
nondeterministic
pushdown multicounter automata
is decidable. \cite{I78}
\end{lemma}
When an automaton does not have an input tape,
we call it a machine. In this case, 
we are interested in
the behaviors generated by the machine
rather than the language accepted by the automaton.
We shall use NPCM (resp. NCM) to stand for
NCPA (resp. NCA) without an input tape.

Let $\N$ be integers, 
$\F=\Q$ (rationals) or $\R$ (reals), 
 $\Gamma$ be an alphabet.
We use $\N^+$ and $\D$ to denote non-negative
values in $\N$ and $\F$, respectively.
Each value $v\in \D$ can be uniquely
expressed as the sum
of $\up v + \down v$,
where $\up v\in \N$ is the integral part of $v$, and
$0\le \down v<1$ is the fractional part of $v$.
A {\em dense variable}
is a variable over $\F$.
An {\em integer  variable}
is a variable over $\N$.
A {\em word  variable}
is a variable over $\Gamma^*$.
Let $m\ge 1$.
For each $1\le i\le m$,
we use $x_i$, $y_i$, and $w_i$ to denote
a dense variable, an integer variable, and
a word variable,  respectively.
We use $\#_a(w_i)$ to denote a {\em count variable}
 representing 
the number of symbol $a\in\Gamma$ in $w_i$.
A {\em linear term} $t$ is defined as follows:
$$t::= ~n ~~|~~ x_i ~~|~~ y_i ~~|~~ \#_a(w_i) ~~|~~ t-t ~~|~~ t+t,$$
where $n\in \N, a\in \Gamma$ and $1\le i\le m$. 
A {\em mixed linear relation} $l$ is defined as follows:
$$l::=~ t>0 ~~|~~  t=0 ~~|~~  t_{discrete} ~mod~ n=0 ~~|~~  \neg l 
~~|~~  l \land l,$$
where $t$ is a linear term,
$0\ne n\in \N$,
 and $t_{discrete}$ is a linear
term
not containing dense variables.
Notice that a mixed linear relation could contain
dense variables, integer variables and
word count variables.
A {\em dense linear relation} is a mixed linear relation
that contains dense variables only.
A {\em discrete linear relation} is a mixed linear relation
that does not
contain dense variables.
Obviously,
any discrete linear relation is a Presburger 
formula over integer variables and
word count variables.

Each integer can be represented as
 a unary string, e.g.,
string ``$00000$" (resp. ``$11111$") for integer $+5$ (resp. $-5$).
In this way, a tuple of integers and words can be encoded
as a string by concatenating the unary representations of each 
integer and each of the  words, with a separator $\#\not\in\Gamma$.
For instance, $(2, -4, w)$ is encoded as string ``$00\#1111\#w$".
Consider a predicate $H$ over integer variables and word variables.
The domain of $H$ is the set of
tuples
of integers and words that satisfy $H$.
Under the encoding, the domain  of $H$ 
can be
treated as  a set of strings, i.e., a language.
A predicate $H$ over integer variables and word variables
 is an {\em NPCA predicate} (or simply NPCA)
if
there is an NPCA accepting the domain of $H$.
A {\em $\FNPCM$-formula} $f$ is defined as follows:
$$f::= ~ l_{dense}\land H ~~|~~  l_{dense}\lor H ~~|~~  f\lor f,$$
where $l_{dense}$ is a dense linear relation and
$H$ is an NPCA predicate.
Therefore, a $\FNPCM$ formula is a finite disjunction of
formulas in the form of $l_{dense}\land H$ or
$l_{dense}\lor H$, where dense variables (contained only
in each $l_{dense}$) and discrete variables (contained only
in each $H$) are separated.
Let $p, q, r\ge 0$.
A predicate $A$ on tuples in $\F^p\times \N^q\times (\Gamma^*)^r$
is {\em $\FNPCM$-definable}
 if there is a $\FNPCM$-formula $f$ with
$p$ dense variables, $p+q$ integer variables, and
$r$ word variables, such that, for all
$x_1,\cdots, x_p$ in $\F$, for 
all $y_1,\cdots, y_q$ in $\N$, and for all
$w_1,\cdots, w_r$ in $\Gamma^*$,
$$(x_1,\cdots, x_p, y_1,\cdots, y_q, w_1,\cdots, w_r)\in A
$$
$$\mbox{~iff~}
f(\down{x_1}, \cdots, \down{x_p},
\up{x_1}, \cdots, \up{x_p}, y_1,\cdots, y_q, w_1,\cdots, w_r)
\mbox{~holds.}
$$
\begin{lemma}\label{basicprop}
(1). Both
 $l_{discrete}\land H$ and $l_{discrete}\lor H$ are
NPCA predicates, if $l_{discrete}$ is a discrete linear relation
and $H$ is an NPCA predicate.

(2). NPCA predicates are closed under existential quantifications
(over integer variables and word variables).

(3). If $A$ is $\FNPCM$-definable and $l$ is a mixed linear relation,
then both $l\land A$ and $l\lor A$ are $\FNPCM$-definable.

(4). The emptiness (or satisfiability) problem for 
 $\FNPCM$-definable predicates is decidable.
\end{lemma}
\begin{proof}
(1). $l_{discrete}$ is  a Presburger formula.
(The domain of) $l_{discrete}$ can therefore
be accepted by a deterministic NCA \cite{I78}.
Hence, $l_{discrete}\land H$ and $l_{discrete}\lor H$
can be accepted by NPCAs by ``intersecting" and
``joining" the deterministic NCA and the NPCA that accepts $H$,
respectively.

(2). Let $H$ be an NPCA predicate containing variable
$z$ (either an integer variable or a word variable).
Assume $H$ is accepted by NPCA $M$.
An NPCA $M'$ can be constructed to
accept $\exists z H$ by guessing each symbol
in the encoding of $z$ (on the input tape of $M$)
and simulating $M$.

(3).
We first show that any mixed linear relation
$l$ is definable by a {\em separately}
mixed linear relation $l'$ (i.e., $l'$ is a Boolean combination
of dense linear relations and discrete linear relations. So,
$l'$ does not have a term containing both
dense variables and discrete variables.).
That is, for all 
$x_1,\cdots, x_p\in \F, y_1,\cdots, y_q\in \N$, 
$$l(x_1,\cdots, x_p,y_1,\cdots, y_q)\mbox{~iff~}
l'(\down{x_1}, \cdots,\down{x_p},
\up{x_1},\cdots, \up{x_p}, y_1,\cdots, y_q).$$
Instead of giving a lengthy proof, we look at an example of $l$:
$x_1-x_2+y_1>2$. This can be rewritten  as:
$\up{x_1}-\up{x_2}+y_1-2+\down{x_1}-\down{x_2}>0.$
Term $\down{x_1}-\down{x_2}$ is the only part containing
dense variables. Since $\down{x_1}-\down{x_2}$ is bounded,
separating cases for this term
 being at (and between) -1, 0, 1 will
give a separately mixed linear relation $l'$.
This separation idea can be applied for any mixed linear relation $l$.
If $A$ is definable by a
$\FNPCM$-formula $f$,
then $l\land A$ (resp. $l\lor A$) is definable by
$l'\land f$
(resp. $l'\lor f$). By re-organizing the
dense linear relations
(in $l'$ and $f$)
 and
the discrete linear relations (in $l'$)
 such that the discrete linear relations
are grouped with the NPCA predicates in $f$,
$l'\land f$ and $l'\lor f$ can be made
$\FNPCM$-formulas using
Lemma \ref{basicprop} (1).

(4). The emptiness problem 
for $l_{dense}\land H$ and
$l_{dense}\lor H$ is decidable, noticing that
the emptiness for $l_{dense}$, which is
expressible in the additive theory of
reals (or rationals), is decidable,
and
the emptiness of NPCA predicate $H$ is decidable (Lemma \ref{emptiness}).
Therefore, the emptiness of any
$\FNPCM$ formulas, as well as, from  Lemma \ref{basicprop} (3), any
$\FNPCM$-definable predicates,
 is decidable.
\qed
\end{proof}

\section{Clock Patterns and Their Changes
}\label{pattern}

A dense clock is simply a dense variable
taking non-negative values in $\D$.
Now we fix a $k> 0$ and consider $k+1$ clocks
$\x=x_0, \cdots, x_k$. For technical reasons, $x_0$ is an auxiliary
clock indicating the current time $now$.
 Let $K=\{0,\cdots,k\}$, and
$K^+=\{1,\cdots,k\}$. 
A subset $K'$ of $K$ is abused as
a set of clocks; i.e., we say $x_i\in K'$ if $i\in K'$.
A {\em (clock) valuation} $\v$ is a function
$K\to \D$ that assigns a  value in $\D$ to
each clock in $K$.
A {\em discrete (clock) valuation} $\u$ is a function
$K\to \N^+$ that  assigns a  value in $\N^+$
 to 
each clock in $K$. 
For each valuation $\v$ and $\delta\in \D$,
$\up \v$, $\down \v$ and $\v+\delta$
are valuations satisfying
$\up \v(i)=\up{\v(i)}$, $\down \v(i)=\down{\v(i)}$
and $(\v+\delta)(i)=\v(i)+\delta$ for each $i\in K$.
The {\em relative representation}
$\relv \v$ of a valuation $\v$ is a  valuation satisfying:
\begin{itemize}
\item $\up{\relv \v}=\up \v$,
\item $\down{\relv \v}(0)=\down{1-\down{\v}(0)}$,
\item $\down{\relv \v}(i)=\down{\down{\v}(i)+\down{\relv \v}(0)}$, 
for each $i\in K^+$.
\end{itemize}
A valuation $\v_0$ is {\em initial } if 
the auxiliary clock $x_0$ has value 0 in $\v_0$.

\begin{example}\label{ex1}
Let $k=4$ and $\v_1=(4.296, 1.732, 1.414, 5.289, 3.732)$.
It can be calculated that
$\relv \v_1=
(4.704, 1.436, 1.118, 5.993, 3.436).$
Let   $\v_2=\v_1+.268=(4.564, $ $2,$ $ 1.682, 5.557, 4)$.
Then, $\relv \v_2=
(4.436, 2.436, 1.118, 5.993, 4.436).$
It is noticed that
all the fractional parts (except for $\relv{\v_1}(0)$ and $\relv{\v_2}(0)$)
are the same in $\relv \v_1$ and $\relv \v_2$.
It is easy to show that
a clock progress (i.e.,
$x_0,\cdots,
x_k$ progress by the same amount such as .268)
 will not  change the
fractional parts of clock values (for clocks $x_1,\cdots,x_k$) in
a relative representation.
\hfill\qed
\end{example}

\subsection{Clock Patterns}

\begin{figure*}[tbph]
\centerline{\epsfig{file=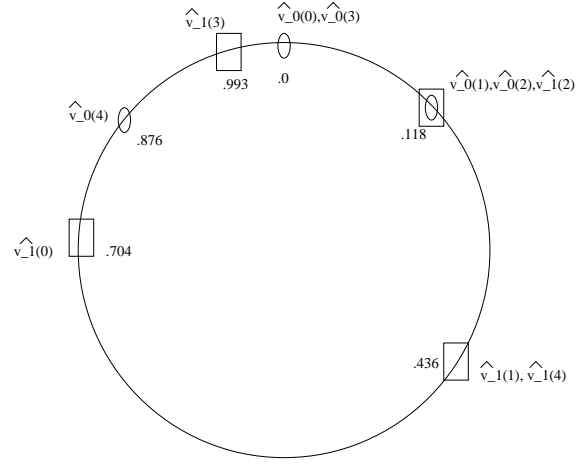,width=3.0in, height=2.4in}}
\caption{A graphical representation of the fractional parts of
$\v_0$ and $\v_1$ in a relative representation
in Example \ref{ex2}. That is, $\down{\relv{\v_0}}=(0, .118, .118, 0, .876)$
and $\down{\relv{\v_1}}=(.704, .436, .118, .993, .436)$ as
in Example \ref{ex2} and \ref{ex1}. Ovals are for components in
$\down{\relv{\v_0}}$  and boxes  are for components in
$\down{\relv{\v_1}}$. For instance, the oval labeled by $\relv{\v_0}(4)$
corresponds to $\down{\relv{\v_0}}(4)=.876$.
} \label{fig1}
\end{figure*}

We distinguish two disjoint 
 sets, $K^0=\{0^0,\cdots,k^0\}$
and $K^1=\{0^1,\cdots,k^1\}$, of indices.
A {\em pattern} $\eta$ is
a sequence $$p_0,\cdots,p_n,$$
 for some $0\le n< 2(k+1)$,
of nonempty and disjoint subsets of $K^0\cup K^1$ such that
\begin{itemize}
\item $0^0\in p_0$ and
\item $\cup_{0\le i\le n}p_i=K^0\cup K^1$. 
\end{itemize}
In pattern $\eta$, $p_i$ is called the {\em $i$-position}.
A pair of valuations $(\v_0,\v_1)$ is {\em initialized}
if $\v_0$ is initial.
The pattern of $(\v_0,\v_1)$ characterizes the fractional ordering
between elements in  $\down{\relv{\v_0}}$ and $\down{\relv{\v_1}}$
(where  $K^0$ is for indices of $\v_0$ and
$K^1$ is for indices of $\v_1$). Formally,
an initialized pair
$(\v_0,\v_1)$ {\em has pattern $\eta=p_0,\cdots,p_n$}, written
$(\v_0,\v_1)\in \eta$, or $[(\v_0,\v_1)]=\eta$, if,
for each $0\le m, m'\le n$,
each $b, b'\in \{0,1\}$,
and
each $i,i'\in K$,
$i^{b}\in p_m$  and ${i'}^{b'}\in p_{m'}$ imply
that
$$\down{\relv{\v_{b}}}(i)=\down{\relv{\v_{b'}}}(i') ~(\mbox{resp.}~ <) 
\mbox{~iff~}
m=m' ~(\mbox{resp.}~ m<m').$$
Though this definition of a pattern is quite complex,
a pattern can be easily visualized
after looking at the following example.
\begin{example}\label{ex2}
Consider $\v_1$ in  Example \ref{ex1} and
an initial valuation
$\v_0=(0, 3.118, 5.118,$ $ 2, 1.876)$. Since $\v_0$ is initial,
$\relv{\v_0}=\v_0$. The fractional parts of $\v_0$ and $\v_1$,
in the relative representation, can be
put on a big circle representing the interval $[0,1)$ as shown in
Figure \ref{fig1}. Each
 fractional value  $\down{\relv{\v_0}}(i)$ for $\v_0$ is
represented by an oval;
each  fractional value  $\down{\relv{\v_1}}(i)$ for $\v_1$ is
represented by a box.
The pattern of $(\v_0,\v_1)$ can be drawn
by collecting clockwisely (from the top, i.e., $\relv{\v_0}(0)= 0$)
 the indices (superscripted with 0, e.g., $3^0$ for
$\relv{\v_0}(3)$) for each component in
$\relv{\v_0}$ and  the indices (superscripted with 1, e.g., $3^1$ for
$\relv{\v_1}(3)$) for each component in
$\relv{\v_1}$; i.e., the pattern is
$$\eta=p_0,\cdots,p_5$$
with
$p_0=\{0^0,3^0\}, p_1=\{1^0,2^0,2^1\},
p_2=\{1^1,4^1\}, p_3=\{0^1\}, p_4=\{4^0\}, p_5=\{3^1\}.$
\qed
\end{example}

There are at most $2^{6(k+1)^2}$ distinct
 patterns. Let $\Phi$ 
denote the set of all the patterns (for the fixed $k$).
A pattern is {\em initial} if it is the pattern of
$(\v_0,\v_0)$ for some initial valuation $\v_0$.
If $\eta$ is the pattern of
$(\v_0,\v_1)$, we use $init(\eta)$ to denote the pattern of
$(\v_0,\v_0)$. $init(\eta)$ is unique for each $\eta$.
Given two initialized pairs $(\v_0^1, \v_1)$
and $(\v_0^2, \v_2)$, we write $(\v_0^1, \v_1)\rdeq (\v_0^2, \v_2)$, if
$(\v_0^1, \v_1)$ and $(\v_0^2, \v_2)$ have the same pattern,
and have
the same integral parts (i.e.,
$\up{\v_0^1}=\up{\v_0^2}, \up{\v_1}=\up{\v_2}$).
The following lemma can be observed. 
\begin{lemma}\label{initialpattern}
For any two initialized pairs $(\v_0^1, \v_1)$
and $(\v_0^2, \v_2)$
with $(\v_0^1, \v_1)\rdeq (\v_0^2, \v_2)$,
the following statements hold:

(1). the pattern of $(\v_0^1, \v_1)$ is initial iff
$\down{\v_1}=\down{\v_0^1}$,

(2). $\v_1$ is initial (i.e., $\v_1(0)=0$) iff $\v_2$ is initial,

(3). $\v_1=\v_0^1$ iff $\v_2=\v_0^2$.
\end{lemma}

A valuation
$\v_1$ {\em has pattern} $\eta$ if
there is an initial $\v_0$ such that 
$(\v_0,\v_1)$ has pattern $\eta$. $\v_1$ 
may have a number of patterns, by different choices of $\v_0$.
A pattern of $\v_1$ tells
the truth values of all the {\em fractional orderings}
$\down{\v_1}(i)\#\down{\v_1}(j)$ and $\down{\v_1}(i)\#0$
(where $\#$ stands for $<,>,\le,\ge,=$.), for all $i,j\in K^+$,
as shown in the following lemma.


\begin{lemma}\label{densetest}
Let $\eta=p_0,\cdots,p_n$ be a pattern of a valuation $\v$.
Assume $0^1\in p_i$ for some $0\le i\le n$.
Then, for any 
$m_1$ and $m_2$ (with $0\le m_1,m_2\le n$),
for any $j_1$ and $j_2$ in $K^+$
(with $j_1^1\in p_{m_1}$ and $j_2^1\in p_{m_2}$),
the following 
statements hold.

 (1). $\down{\v}(j_1)>\down{\v}(j_2)$ iff
one of the following conditions holds:

~~~~~~~~~~~~~~$m_1<i\le m_2$, 

~~~~~~~~~~~~~~$m_2<m_1<i$, 

~~~~~~~~~~~~~~$i\le m_2<m_1$.

(2). $\down{\v}(j_1)=\down{\v}(j_2)$ iff
$m_1=m_2$. 

(3). $\down{\v}(j_1)>0$ iff
$m_1\ne i$.

(4). $\down{\v}(j_1)=0$ iff
$m_1=i$.
\end{lemma}
\begin{proof}
Directly from the definition of a pattern.
\qed
\end{proof}

Recall
$0^1\in K^1$
stands for
the index for the value of
clock $x_0$ (representing $now$) in $\v_1$.
Let $\eta=p_0,\cdots,p_n$ be a pattern.
$p_i$ is the {\em now-position} of $\eta$
if $0^1\in p_i$.
A pattern $\eta$
is {\em regulated} if
the now-position of $\eta$ is $p_0$.
Note that the pattern of
an initialized pair $(\v_0,\v_1)$ is regulated if and only if
the auxiliary clock $x_0$ takes an integral value
in $\v_1$ (i.e., $\down{\v_1}(0)=0$).
A pattern is  a {\em merge-pattern} if
the now-position is a singleton set (i.e., $0^1$ is the only element).
A pattern is  a {\em split-pattern} if
it is not a merge-pattern, i.e., the now-position
contains more than one element. (``Merge" and ``split" will be made clear
in a moment.)
Obviously, a
 regulated pattern is always a split-pattern.
This is because the now-position  of
a regulated pattern, which is $p_0$, contains at least
two elements $0^0$ and $0^1$.

\subsection{Clock Progresses}

For each $0< \delta\in \D$,
$\v+\delta$ is the result of a clock progress
 from
$\v$ by an amount of $\delta$.
How does a pattern change according to the progress?
Let us first look at an example.

\begin{example}\label{ex4}
Consider $\v_1,\v_2$ ($=\v_1+.268$) in Example \ref{ex1},
and
$\v_0$ in Example \ref{ex2}.
In Example \ref{ex2}, we indicated that the pattern
$\eta_1$ of $(\v_0,\v_1)$ is
$$\{0^0,3^0\}, \{1^0,2^0,2^1\},
\{1^1,4^1\}, \{0^1\}, \{4^0\},\{3^1\}.$$
Similar steps can be followed
to show that
the pattern
$\eta_2$ of $(\v_0,\v_2)$ is
$$\{0^0,3^0\},\{1^0,2^0,2^1\},
\{1^1,4^1,0^1\}, \{4^0\},\{3^1\}.$$
A helpful way to see the relationship
between $\eta_1$ and $\eta_2$ is by looking at Figure \ref{fig1}.
Holding the box labeled by $\relv{\v_1}(0)$ (for the current time)
and sliding counter-clockwisely along the big circle
for an amount of $.268$ will stop at the box labeled
by $\relv{\v_1}(1)$ and $\relv{\v_1}(4)$.
Thus, the pattern $\eta_2$ (after
sliding) is exactly
 $\eta_1$ (before sliding)
except that $0^1$ in the
$3$-position in $\eta_1$ is merged into
the $2$-position in $\eta_2$.
Notice that $\eta_1$ is a merge-pattern
and
the resulting
$\eta_2$ is a split-pattern.
The integral parts $\up{\v_1}(1)$ and
$\up{\v_1}(4)$ change to
$\up{\v_2}(1)=\up{\v_1}(1)+1$ and $\up{\v_2}(4)=\up{\v_1}(4)+1$.
But all the other components of $\up{\v_1}$ do not change.
The reason is that, after merging
$0^1$ with
$1^1$ and $4^1$ in $\eta_2$, the fractional parts
$\down{\v_2}(1)$ and $\down{\v_2}(4)$ are ``rounded" (i.e.,
become 0).
What if we further make a clock progress from $\v_2$ for an amount of
$\delta'=.12$? The resulting pattern
$\eta_3$ of $(\v_0, \v_3)$ with $\v_3=\v_2+\delta'$
is the result of splitting $0^1$ from the 2-position
$\{1^1, 4^1, 0^1\}$. That is,
$\eta_3$ is
$$\{0^0,3^0\}, \{1^0,2^0,2^1\},
\{0^1\}, \{1^1,4^1\}, \{4^0\},\{3^1\},$$
which is a merge-pattern again.
This process of merging and splitting
can be formally defined as the following function $next$.
\qed
\end{example}

Function $next: \Phi\times (\N^+)^{k+1}\to \Phi\times (\N^+)^{k+1}$
describes how  a pattern changes upon a clock progress.
Given any discrete valuation $\u$ and pattern 
$\eta=p_0,\cdots,p_n$ with 
the now-position being $p_i$
for some $i$,
$next(\eta, \u)$ is defined to be  $(\eta', \u')$ such that,
\begin{itemize}
\item (the case when $\eta$ is a merge-pattern) if $i> 0$ and $|p_i|=1$
(that is, the now-position
$p_i=\{0^1\}$), then
$\eta'$  is $$p_0, \cdots, p_{i-1}\cup\{0^1\}, p_{i+1}, 
\cdots,p_n$$ (that is,
$\eta'$ is the result of merging
the now-position to the previous position), and for each $j\in K^+$, if
$j^1\in p_{i-1}$, then $\u'(j)=\u(j)+1$ else $\u'(j)=\u(j)$.
Besides, if $i=1$ (i.e., the now-position is merged to $p_0$;
in this case, $\eta'$ is a regulated pattern), then 
$\u'(0)=\u(0)+1$  else $\u'(0)=\u(0)$,
\item (the case when $\eta$ 
is a split pattern) 
if $i\ge0$ and $|p_i|>1$, then
$\eta'$ is 
the result of splitting $0^1$ from the now-position.
That is, if $i>0$, $\eta'$ is 
$$p_0, \cdots, p_{i-1}, \{0^1\}, p_{i}-\{0^1\},
p_{i+1},
\cdots,p_n.$$
However, if $i=0$,
$\eta'$ is 
$$p_0-\{0^1\}, p_1, \cdots, p_n, \{0^1\}.$$
In either case, $\u'=\u$.
\end{itemize}
If $next(\eta, \u)=(\eta', \u')$,
(1). $\eta'$
is called {\em the next pattern of $\eta$}, written
$Next(\eta)$, (2). 
$\Delta_{\eta}\in \{0,1\}^{k+1}$ is called the {\em increment vector}
of $\eta$ with $\Delta_{\eta}=\u'-\u$.
Obviously,
$Next(\eta)\ne \eta$ and
$Next(\cdot)$ is total and 1-1.

\begin{figure*}[tbph]
\centerline{\epsfig{file=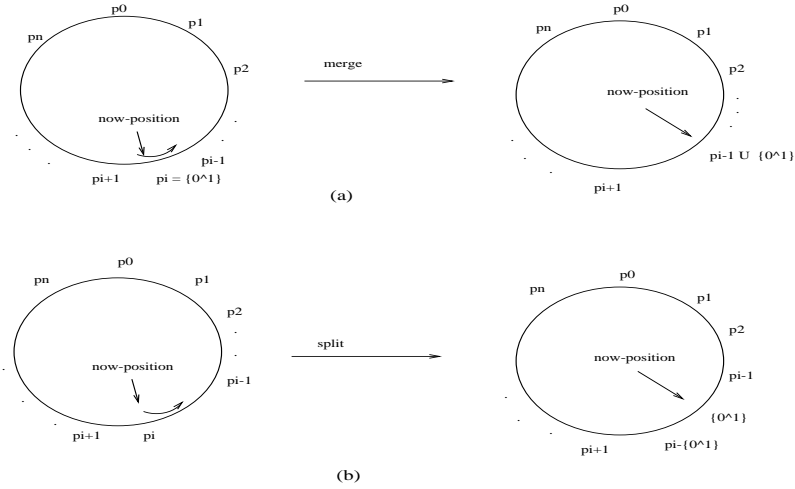,width=4.2in, height=2.5in}}
\caption{A graphical representation of a pattern
$\eta=p_0,\cdots,p_n$.
Operator
$Next(\cdot)$ has the same effect
as moving the now-position
counter-clockwisely.
In case (a), the now-position is merged
to the previous position.
In case (b), 
index $0^1$
is split from the now-position.
} \label{fig3}
\end{figure*}

To better understand $Next(\cdot)$,
we visualize pattern
$\eta$
as a circle shown in Figure \ref{fig3}.
Applications of $Next(\cdot)$ can be regarded as
moving the index $0^1$ along the circle, by performing
merge-operations (Figure \ref{fig3} (a))
and split-operations (Figure \ref{fig3} (b))
alternatively.
After enough number of applications of $Next(\cdot)$,
$0^1$ will return to the original now-position
after moving through the entire circle.
That is,
for each pattern $\eta$,
there is a smallest positive integer $m$ such that
$Next^{m}(\eta)=\eta$; i.e.,
 $\eta_0,\cdots,\eta_m$ satisfies
$\eta_0=\eta_m=\eta$, and $Next(\eta_i)=\eta_{i+1}$
for each $0\le i<m$. 
More precisely,
by looking at Figure \ref{fig3},
if $\eta$ is a merge-pattern, $m=2n$;
if $\eta$ is a split-pattern, $m=2(n+1)$.
Furthermore, elements $\eta_0,\cdots,\eta_{m-1}$ are
distinct.
The sequence $\eta_0,\cdots,\eta_m$ is called a {\em pattern ring}.
The pattern ring is unique for each fixed $\eta_0$. 
Notice that $next^m(\eta, \u)=(\eta, \u+1)$ for each $\u$.
Since the next pattern $Next(\eta)$ is a merge-pattern (resp. split-pattern)
if $\eta$ is a split-pattern (resp. merge-pattern),
on a pattern ring, merge-patterns and split-patterns appear alternately.

Fix any initialized pair $(\v_0,\v)$ and $0<\delta\in\D$.
Assume the patterns of $(\v_0,\v)$ and
$(\v_0,\v+\delta)$ are $\eta$ and $\eta'$, respectively.
We say {\em $\v$ has no pattern change for $\delta$}
if, for all $0\le \delta'\le\delta$,
$(\v_0,\v+\delta')$ has the same pattern.
We say {\em $\v$ has one pattern change for $\delta$}
if $Next(\eta)=\eta'$ 
 and,
for all $0< \delta'<\delta$,
$(\v_0,\v+\delta')$ has pattern $\eta$, or,
for all $0< \delta'<\delta$,
$(\v_0,\v+\delta')$ has pattern $\eta'$.
The following lemma
on the correctness of $next$
can be observed.

\begin{lemma}\label{progresspatternchange}
For any initialized pair
$(\v_0,\v)$ and any
$0<\delta\in\D$,
the following statements are equivalent:

(1). 
$next([(\v_0,\v)],\up{\v})=([(\v_0,\v+\delta)],\up{\v+\delta}),$

(2). 
$\v$ has one pattern change for $\delta$.
\end{lemma}

We say {\em $\v$ has $n$ pattern changes for $\delta$} with $n\ge 1$,
if there are positive $\delta_1,\cdots,\delta_n$
in $\D$ with $\Sigma_{1\le i\le n}\delta_i=\delta$
such that
$\v+\Sigma_{1\le i\le j}\delta_i$ has one pattern change for
$\delta_{j+1}$, for each $j=0,\cdots,n-1$.
It is noticed that
for any $\delta\le 1$,
$\v$ has at most $m$ pattern changes, where
$m$ is the length of the pattern ring starting from
the pattern $\eta$ of 
$(\v_0,\v)$. This $m$ is uniformly bounded by
$4(k+1)$.

\begin{lemma}\label{maxpatternchange}
For any initialized pair
$(\v_0,\v)$ and any
$\delta\in \D$,
(1) $\v$ has at most $4(k+1)$
pattern changes for $\delta$ if $\delta\le 1$,
(2) $\v$ has at least one pattern change for $\delta$  
if $\delta\ge 1$, 
(3) if $\v$ has no pattern change for
$\delta$
 then $\up{\v}=\up{\v+\delta}$. 
\end{lemma}

\subsection{Clock Resets}

In addition to  clock progresses,
clock resets are the other form
of clock behaviors. 
Let $r\subseteq K^+$ be (a set of) {\em clock resets}.
$\v\downarrow_r$ denotes
 the result of resetting each clock $x_i\in r$ (i.e.,
$i\in r$).
That is, for each $i\in K$,
\[ (\v\downarrow_r)(i)=\left\{\begin{array}{ll}
0& \mbox{if $i\in r$}\\
\v(i)  & \mbox{otherwise.}
\end{array}
\right. \]

\begin{example}\label{ex6}
Consider $\v_0$ and $\v_1$ given
in Example \ref{ex2} and
Example \ref{ex1}.
Assume $r=\{4\}$. By definition,
$\v_1\downarrow_r=(4.296, 1.732,$ $ 1.414, 5.289, 0)$.
It can be calculated that the relative representation
of $\v_1\downarrow_r$ is
$(4.704,$ $ 1.436,$ $1.118, 5.993, 0.704)$.
The pattern of $(\v_0, \v_1\downarrow_r)$ can be figured out
again by looking at Figure \ref{fig1}.
The  reset of clock $x_4$ can be conceptually
regarded as moving the label $\relv{\v_1}(4)$ from the box of
$\relv{\v_1}(1)$ and $\relv{\v_1}(4)$ to the box of $\relv{\v_1}(0)$
(the current time).
Therefore, the pattern after the reset changes
from $$\{0^0,3^0\}, \{1^0,2^0,2^1\},
\{1^1,4^1\}, \{0^1\}, \{4^0\},\{3^1\}$$ of $(\v_0,\v_1)$ to
$$\{0^0,3^0\}, \{1^0,2^0,2^1\},
\{1^1\}, \{0^1, 4^1\}, \{4^0\},\{3^1\}$$ of
$(\v_0, \v_1\downarrow_r)$ by
moving $4^1$ into the position containing $0^1$.
\qed
\end{example}

Functions
$reset_r: \Phi\times (\N^+)^{k+1}\to \Phi\times (\N^+)^{k+1}$
for $r\subseteq K^+$
describe how  a pattern changes after clock resets.
Given any discrete valuation $\u$ and any pattern
$\eta=p_0,\cdots,p_n$ with
the now-position being $p_i$
for some $i$,
$reset_r(\eta, \u)$ is defined to be  $(\eta', \u')$ such that,
\begin{itemize}
\item $\eta'$  is $p_0-r^1, \cdots, p_{i-1}-r^1, p_i\cup r^1,
p_{i+1}-r^1, \cdots, p_n-r^1$,
where $r^1=\{j^1: j\in r\}\subseteq K^1$.
Therefore, $\eta'$ is the result of bringing every index in $r^1$
into the now-position. Notice that some of
$p_m-r^1$ may be empty after moving indices in $r^1$ out of $p_m$, for
$m\ne i$. In this case, these
empty elements are removed from $\eta'$ (to guarantee that $\eta'$ is 
well defined.),
\item  for each $j\in K$, if
$j\in r$, then $\u'(j)=0$ else $\u'(j)=\u(j)$.
\end{itemize}
If $reset_r(\eta, \u)=(\eta', \u')$,
$\eta'$
is  written as
$Reset_r(\eta)$.
Note that $Reset_r(\eta)$ is unique for each $\eta$
and $r$,  and is independent
of  $\u$.
The following lemma states that {\em reset} is correct.

\begin{lemma}\label{resetpatternchange}
For any
initialized pair $(\v_0,\v)$
and any $r\subseteq K^+$,
$$reset_r([(\v_0,\v)],\up{\v})=
([(\v_0,\v\downarrow_r)],\up{\v\downarrow_r}).$$
\end{lemma}

\section{Clock Constraints and Patterns}\label{CCP}
An {\em atomic clock constraint} (over clocks
$x_1,\cdots,x_k$, excluding $x_0$)
 is a formula in the form
of $x_i-x_j\# d$ or $x_i\#d$ where
$0\le d\in \N^+$ and $\#$ stands for $<,>,\le,\ge,=$.
A {\em clock constraint}
 $c$ is a Boolean combination of atomic clock constraints.
Let $\C$ be the set of all
clock constraint (over clocks $x_1,\cdots,x_k$).
We say $\v\in c$ if  clock valuation $\v$
 (for $x_0,\cdots,x_k$)
satisfies clock constraint $c$.

Any clock constraint $c$ can be written as
a Boolean combination $I(c)$
 of clock constraints over
discrete clocks $\up{x_1},\cdots,\up{x_k}$
and fractional orderings $\down{x_i}\#\down{x_j}$
and $\down{x_i}\#0$.
For instance,
$x_i-x_j<d$ is equivalent to:
$\up{x_i}-\up{x_j}<d$, or, $\up{x_i}-\up{x_j}=d$
and $\down{x_i}<\down{x_j}$.
$x_i>d$ is equivalent to:
$\up{x_i}>d$, or, $\up{x_i}=d$  and $\down{x_i}>0$.
Therefore, testing $\v\in c$ is equivalent to testing
$\up{\v}$ and the fractional orderings on
$\down{\v}$ satisfying $I(c)$.

Assume $\v$ has a pattern $\eta=p_0,\cdots,p_n$.
A fractional ordering on $\down{\v}$ is equivalent to a
Boolean condition on  $\eta$, as shown in Lemma \ref{densetest}.
Whenever $\eta$ is fixed, each fractional ordering in $I(c)$ has
a specific truth value (either 0 or 1). In this case,
we use $I(c)^\eta$, or simply $c^\eta$, to denote the result of
replacing fractional orderings in $I(c)$ by the truth values given
by $\eta$. $c^\eta$, without containing fractional orderings, is
 just a clock constraint (over discrete clocks).
Notice that the pattern space $\Phi$ is finite,
therefore, $\v\in c$ is equivalent to
$$\bigvee_{\eta\in\Phi} (\v\mbox{~has pattern~}\eta\land \up{\v}\in c^\eta).$$
Hence, the truth value of $\v\in c$ only depends on a
pattern  of $\v$
and the integral parts of $\v$. These observations conclude the following 
results. In particular, Lemma \ref{tests} (2) indicates that
it is sufficient to test the two end points $\v\in c$ and $\v+\delta\in c$
in order to make sure that
 $c$ is consistently satisfied
on each $\v+\delta'$, $0\le \delta'\le\delta$, if 
from $\v$ to $\v+\delta$, there is at most one pattern change.
\begin{lemma}\label{tests}
(1). For any initialized pair $(\v_0,\v)$,
any pattern  $\eta\in\Phi$,
if $(\v_0,\v)$ has pattern $\eta$,
then, for any clock constraint $c\in \C$,
$\v\in c$ iff $\up{\v}\in c^\eta$.

(2). For any initialized pair $(\v_0,\v)$ and
any $0<\delta\in\D$,
if $\v$ has at most one pattern change for $\delta$,
then, for any clock constraint $c\in \C$,
$$\forall 0\le\delta'\le\delta (\v+\delta'\in c)
\mbox{~iff~} \v\in c\mbox{~and~} \v+\delta\in c.$$

(3). 
For any initialized pairs $(\v^1_0, \v_1)$ and
$(\v^2_0, \v_2)$, if
 $(\v^1_0, \v_1)\rdeq (\v^2_0, \v_2)$,
then, for any $c\in \C$,
 $\v_1\in c$ iff $\v_2\in c$.
\end{lemma}
\begin{proof}
(1) is from the observations made before this lemma
in this section.
(2) is from (1) and
Lemma \ref{progresspatternchange}.
(3) is directly from (1).
\qed
\end{proof}

Now, we consider two initialized pairs $(\v_0^1,\v_1)$ and $(\v_0^2,\v_2)$
such that $$(\v_0^1,\v_1)\rdeq (\v_0^2,\v_2).$$
That is,
from the definition of $\rdeq$,
$\v_0^1$ (resp. $\v_1$) has the same 
integral parts as $\v_0^2$ (resp. $\v_2$).
Besides, the two pairs have the same pattern.
From Lemma \ref{tests}(3), any test $c\in \C$
will not tell the difference between $\v_1$ and $\v_2$.
Assume $\v_1$ can be reached from
a valuation $\v^1$ via a clock progress by an amount of $\delta_1$,
i.e., $\v^1+\delta_1=\v_1$. We would like to know whether
$\v_2$ can be reached from
some valuation $\v^2$ also via a clock progress but probably
by a slightly different amount of $\delta_2$ such that
$(\v_0^1,\v^1)$ and $(\v_0^2,\v^2)$ are still equivalent($\rdeq$).
We also expect that for any test $c$, if during
the progress of $\v^1$, $c$ is consistently satisfied, then
so is $c$ for  the progress of $\v^2$.
The following lemma concludes that these, as well as the
 parallel case for clock resets,
can be done. This result can be used later
to show that if $\v_1$ is reached from $\v_0^1$ by a sequence of
transitions that  
repeatedly perform clock progresses and clock resets,
then $\v_2$ can be also reached from $\v_0^2$ via a very similar
sequence such that no test $c$ can distinguish
the two sequences.
 
\begin{lemma}\label{backwards}
For any initialized pairs $(\v_0^1,\v_1)$ and $(\v_0^2,\v_2)$
with  $(\v_0^1,\v_1)$ $\rdeq (\v_0^2,\v_2)$,

(1). for any $0\le \delta_1\in \D$,
for any clock valuation $\v^1$, if $\v^1+\delta_1=\v_1$, then
there exist $0\le \delta_2\in \D$ and clock valuation
$\v^2$ such that

(1.1). $\v^2+\delta_2=\v_2$ and
$(\v_0^1,\v^1)\rdeq (\v_0^2,\v^2)$,

(1.2). $\v^1$ is initial iff $\v^2$ is initial,

~~~~~~~~$\v^1=\v_0^1$ iff $\v^2=\v_0^2$,
and  

~~~~~~~~for any $c\in \C$, $\v^1\in c$ (resp. $\v_1\in c$)
 iff
$\v^2\in c$ (resp. $\v_2\in c$),

(1.3). for any clock constraint $c\in \C$,
$\forall 0\le \delta'\le \delta_1 (\v^1+\delta\in c)$ iff
$\forall 0\le \delta'\le \delta_2(\v^2+\delta\in c)$. 

(2). for any $r\subseteq K^+$, for any clock valuation $\v^1$, if
$\v^1\downarrow_r=\v_1$, then
there exists a valuation
$\v^2$ such that 

(2.1). $\v^2\downarrow_r=\v_2$ and
$(\v_0^1,\v^1)\rdeq (\v_0^2,\v^2)$,

(2.2). same as (1.2).
\end{lemma}
\begin{proof}
(1).
Assume $\delta_1$ is ``small", i.e., from $\v^1$ to $\v_1=\v^1+\delta_1$,
there is at most one pattern change.
Let $\eta=p_0,\cdots,p_n$ be the pattern for $(\v_0^2,\v_2)$
(and, hence,  for $(\v_0^1,\v_1)$).
Assume $0^1\in p_i$ for some $i$.
If $\delta_1$ causes no pattern change for $\v^1$, then simply take
$\delta_2=0$.
If $\delta_1$ causes one pattern change for $\v^1$, then
we put $(\v_0^2,\v_2)$ on a circle (e.g. Figure \ref{fig1}).
If $\eta$ is a split-pattern
(i.e., $|p_i|>1$),
then we separate a new box (only labeled by
$\down{\relv{\v_2}}(0)$) from the original box
labeled by  $\down{\relv{\v_2}}(0)$
and slide the new box backwards (i.e., clockwisely)
for a small positive amount (taken as $\delta_2$)
 without hitting any box or oval.
If $\eta$ is a merge-pattern
(i.e., $|p_i|=1$), then  we slide the box
labeled by $\down{\relv{\v_2}}(0)$
(this is the only label) backwards (i.e., clockwisely)
for a positive amount (taken as $\delta_2$)
until a box or an oval is hit.
Take $\v^2=\v_2-\delta_2$. Obviously,
$(\v_0^1,\v^1)\rdeq (\v_0^2,\v^2)$.
It can be checked that 
(1.2) and
 (1.3) hold
using Lemma \ref{tests} and Lemma \ref{initialpattern}.

Any larger $\delta_1$ that causes
 multiple pattern changes for $\v^1$
can be split
into a finite
(Lemma \ref{maxpatternchange})
 sequence of
small $\delta$'s
that
causes exactly one pattern change.
In this case, $\delta_2$ can be calculated by
working on each small $\delta$ (the last one first)
as in the above proof.

(2). The case when $r=\emptyset$ is obvious.
Assume $r$ contains only
one element $j\in K^+$.
 Assume $\eta$ is the pattern of $(\v^1_0,\v^1)$.
A desired $\v^2$ is picked as follows.
The integral parts of $\v^2$ are exactly
those of $\v^1$; i.e.,
$\up{\v^2}=\up{\v^1}$.
The fractional parts  of $\v^2$ are 
exactly
those of $\v_2$,
except that, in the relative
representation,
$\down{\relv{\v^2}}(j)$ may be different from
$\down{\relv{\v_2}}(j)$. 
Then what is $\down{\relv{\v^2}}(j)$?
It is chosen such that
the pattern of $(\v_0^2,\v^2)$ is $\eta$.
For instance, if $\down{\relv{\v^1}}(j)$ equals to, say,
$\down{\relv{\v_1}}(j_1)$ (resp. $\down{\relv{\v_0^1}}(j_1)$), for some
$j_1$, then
$\down{\relv{\v^2}}(j)$ is picked as
$\down{\relv{\v_2}}(j_1)$ (resp. $\down{\relv{\v_0^2}}(j_1)$).
If $\down{\relv{\v^1}}(j)$ lies strictly between,
say,
$\down{\relv{\v_1}}(j_1)$ (or, $\down{\relv{\v_0^1}}(j_1)$)
and $\down{\relv{\v_1}}(j_2)$ (or, $\down{\relv{\v_0^1}}(j_2)$),
for some $j_1$ and $j_2$, such that
no other component in $\down{\relv{\v^1}}$ and $\down{\relv{\v_0^1}}$
lies strictly between these two values,
then $\down{\relv{\v^2}}(j)$ is picked as any value
lies strictly between
$\down{\relv{\v_2}}(j_1)$ (or, $\down{\relv{\v_0^2}}(j_1)$)
and $\down{\relv{\v_2}}(j_2)$ (or, $\down{\relv{\v_0^2}}(j_2)$)
accordingly.
Since $(\v^1_0,\v_1)\rdeq(\v^2_0,\v_2)$, we can show
$\down{\relv{\v^2}}(j)$ can always be picked.
The choice of $\down{\relv{\v^2}}(j)$ guarantees that
the pattern of $(\v^1_0,\v^1)$ is the same as the pattern
of $(\v^2_0,\v^2)$.
The rest of conditions in (2) can be checked easily.

For the case when $r$ contains more than one element,
the above proof can be generalized by
resetting clocks in $r$ one by one.
\qed
\end{proof}

\section{Pushdown Timed Automata}\label{PTA}
A {\em pushdown timed automaton} (PTA)  $\A$
is 
a tuple
$$\langle S, \{x_1,\cdots,x_k\}, Inv, R, \Gamma,PD\rangle,$$
where
\begin{itemize} 
\item $S$ is a finite set of {\em states},
\item $x_1,\cdots,x_k$ are (dense) clocks,
\item $Inv: S\to \C$ assigns a clock constraint
over  clocks
$x_1,\cdots,x_k$, called
an {\em invariant}, to each state,
\item $R: S\times S\to \C\times 2^{\{x_1,\cdots,x_k\}}$
assigns a clock constraint over  clocks
$x_1,\cdots,x_k$,  called
a {\em reset condition}, and a subset of clocks,
called
clock resets, to a (directed) edge in $S\times S$,
\item $\Gamma$ is the {\em stack alphabet}.
$PD: S\times S\to \Gamma\times\Gamma^*$
assigns a pair $(a,\gamma)$ with $a\in\Gamma$ and
$\gamma\in\Gamma^*$,
called
a {\em stack operation}, to each edge in $S\times S$.
A stack operation $(a,\gamma)$ replaces the top symbol $a$ of the
stack with a string (possibly empty) in $\Gamma^*$.
\end{itemize} 
A {\em timed automaton}
 is a PTA without the pushdown stack.
 
The semantics of $\A$ is defined as follows.
A {\em configuration} is a triple   $(s,\v, w)$  of
a state $s$, a clock valuation $\v$ on $x_0,\cdots,x_k$
(where $x_0$ is the auxiliary clock), and a stack word $w\in\Gamma^*$.
$(s_1,\v_1, w_1)\to_\A (s_2,\v_2, w_2)$ denotes a   {\em 
one-step transition}
of $\A$
 if one 
of the following conditions
 is 
satisfied:
\begin{itemize}
\item ({\em a progress transition})
$s_1=s_2$, $w_1=w_2$,
 and $\exists 0<\delta\in \D$, $\v_2=\v_1+\delta$ and
for all $\delta'$ satisfying $0\le \delta'\le\delta$,
$\v_1+\delta'\in Inv(s_1)$. That is, a 
progress transition makes all the clocks
 synchronously progress 
by amount $\delta>0$, during which the invariant is
consistently satisfied, while the state and the stack content remain unchanged.
\item ({\em a reset transition})
$\v_1\in Inv(s_1)\land c$, $\v_1\downarrow_r=\v_2\in Inv(s_2)$,
and $w_1=aw, w_2=\gamma w$ for some $w\in\Gamma^*$,
where
$R(s_1,s_2)=(c,r)$ for some clock constraint $c$ and clock resets $r$,
and $PD(s_1,s_2)=(a,\gamma)$ for some stack symbol $a\in\Gamma$ and
string $\gamma\in\Gamma^*$.
That is, a reset transition, by moving
 from state $s_1$ to state $s_2$,
 resets
 every clock in $r$ to 0 and keeps all the other clocks unchanged.
The stack content is modified according to the stack operation
$(a,\gamma)$ given  on 
edge $(s_1,s_2)$. Clock values before the transition satisfy
the invariant $Inv(s_1)$ and the reset condition $c$;
clock values after the transition satisfy
the invariant $Inv(s_2)$.
\footnote{
A reader might wonder why we don't have a stack operation 
for a progress transition. That is, a state
$s$ can also be assigned with
a stack operation $(a,\gamma)$ such that each progress transition
by an amount $\delta>0$
on state $s$ also modifies the stack content according to $(a,\gamma)$.
However, this progress transition
can be treated as a sequence of three transitions:
 a progress transition (without a stack operation) 
by $\delta_1>0$, a clock reset transition (by adding a dummy clock)
performing stack operation $(a,\gamma)$,
followed by
a progress transition (without a stack operation) 
 by $\delta_2>0$, whenever $\delta=\delta_1+\delta_2$.
A translation can be worked out by expressing any
PTA
with a stack operation
for each progress transition by
 a PTA defined in this paper.
Since we focus on the clock/stack behaviors
of a PTA,
instead of
the $\omega$-language accepted by it,
input symbols are not considered in our definition.
(The input to a timed automaton is always one-way.
Thus,
input symbols can always be built into states.)
}
\end{itemize}
We write $\to^*_\A$ to be the
 transitive
closure of $\to_\A$. 
Given two valuations $\v^1_0$ and $\v_1$,  two states
$s_0$ and $s_1$, and two stack words 
$w_0$ and $w_1$,
 assume the auxiliary clock $x_0$ starts from 
0, i.e., $\v^1_0$ is initial.
The following result is surprising.
It states that, for {\bf any}
initialized pair
$(\v^2_0, \v_2)$ with $(\v^1_0, \v_1)\rdeq (\v^2_0, \v_2)$,
$(s_0,\v^1_0, w_0)\to^*_\A (s_1,\v_1,w_1)$ 
if and only if
$(s_0,\v^2_0, w_0)$ $\to^*_\A (s_1,\v_2, w_1).$
This result implies that, from the definition of 
$\rdeq$, for any fixed $s_0, s_1, w_0$ and $w_1$,
the pattern of $(\down{\v^1_0}, \down{\v_1})$
(instead of the actual values of $\down{\v^1_0}$ and $\down{\v_1}$), 
the integral values
$\up{\v^1_0}$, and the integral values
$\up{\v_1}$ are sufficient
 to
 determine whether $(s_0,\v^1_0, w_0)$ can reach
$(s_1,\v_1, w_1)$ in $\A$. 
\begin{lemma}\label{strongbisim}
Let $\A$ be a PTA.
 For any states $s_0$ and $s_1$,
any two initial clock valuations $\v^1_0$ and $\v^2_0$,
any two clock valuations $\v_1$ and $\v_2$, and any two stack words $w_0$ and
$w_1$,
if 
$(\v^1_0, \v_1)\rdeq (\v^2_0, \v_2)$, then,
$$(s_0,\v^1_0, w_0)\to^*_\A (s_1,\v_1, w_1)\mbox{~iff~}
(s_0,\v^2_0, w_0)\to^*_\A (s_1,\v_2, w_1).$$
\end{lemma}
\begin{proof}
Lemma \ref{backwards} and
Lemma \ref{tests} already give the result, but for
$\to_\A$ instead of $\to^*_\A$, noticing that
Lemma \ref{tests} guarantees that tests (and obviously stack operations)
are consistent in $(s_0,\v^1_0, w_0)\to_\A (s_1,\v_1, w_1)$
and in $(s_0,\v^2_0, w_0)\to_\A (s_1,\v_2, w_1).$
An induction (on the length of $\to^*_\A$)
can be used to show the lemma, by working from
$(s_1,\v_1, w_1)$ back to $(s_0,\v^1_0, w_0)$.
\qed
\end{proof}

\begin{example}\label{ex100}
It is the time to show
an example to convince the reader that
Lemma \ref{strongbisim} indeed works.
Consider a timed automaton $\A$ shown in Figure \ref{fig2}.
\begin{figure*}[tbph]
\centerline{\epsfig{file=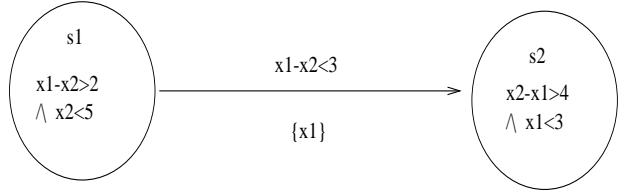,width=3.2in, height=1.0in}}
\caption{An example timed automaton $\A$.}\label{fig2}
\end{figure*}
Let $\v_0^1=(0, 4.98, 2.52),$ $\v_3^1=(5.36, 2.89, 7.88)$.
$(s_1, \v_0^1)\to^*_\A (s_2, \v_3^1)$
is witnessed by:
$(s_1, \v_0^1)\to_\A$ (progress by $2.47$ at $s_1$)
$(s_1, \v^1_1)\to_\A$ (reset $x_1$ and transit to $s_2$)
$(s_2, \v^1_2)\to_\A$ (progress by $2.89$ at $s_2$)
$(s_2, \v_3^1).$
Take a new pair
$\v_0^2=(0, 4.89, 2.11),$ $\v_3^2=(5.28, 2.77, 7.39).$
It is easy to check $(\v_0^1, \v_3^1)\rdeq (\v_0^2, \v_3^2)$.
From Lemma \ref{strongbisim}, $(s_1, \v_0^2)\to^*_\A (s_2, \v_3^2)$.
Indeed, this is witnessed by
$(s_1, \v_0^2)\to_\A$ (progress by $2.51$ at $s_1$)
$(s_1, \v^2_1)\to_\A$ (reset $x_1$ and transit to $s_2$)
$(s_2, \v^2_2)\to_\A$ (progress by $2.77$ at $s_2$)
$(s_2, \v_3^2).$
These two witnesses differ slightly ($2.47$ and $2.89$, vs.
$2.51$ and $2.77$).
We choose $2.77$ and $2.51$
by looking at the first witness backwardly.
That is,
$\v^2_2$ is picked such that
$(\v_0^2, \v^2_2) \rdeq (\v_0^1, \v^1_2)$.
Then, $\v^2_1$ is picked such that
$(\v_0^2, \v^2_1) \rdeq (\v_0^1, \v^1_1)$.
The existence of $\v^2_2$ and $\v^2_1$
is
guaranteed by
Lemma \ref{backwards}.
Finally,
according to Lemma \ref{backwards} again,
 $\v^2_1$ is able to  go back to
 $\v_0^2$. This is because
$\v^1_1$ goes back to $\v_0^1$ through
 a one-step
transition and $\v_0^1$ is initial.
\hfill\qed
\end{example}

Now, we express $\to^*_\A$ in a form  treating
the integral parts and the
fractional parts of clock values separately.
For any
 pattern $\eta\in\Phi$,  any discrete valuations
$\u_0$ and $\u_1$, and any stack words $w_0$ and $w_1$,
define $(s_0,\u_0, w_0)\to^*_{\A,\eta} (s_1,\u_1, w_1)$
to be 
$$\exists \v_0 \exists \v_1 (
\v_0(0)=0 \land
\up{\v_0}=\u_0 \land
\up{\v_1}=\u_1$$
$$ \land
(\v_0,\v_1)\in\eta \land
(s_0, \v_0, w_0)\to^*_{\A} (s_1,\v_1, w_1)).$$

\begin{lemma}\label{character}
Let $\A$ be a PTA.
For any states $s_0$ and $s_1$,
any initialized pair
$(\v_0,\v_1)$,
and any stack words $w_0$ and
$w_1$,
$(s_0, \v_0, w_0)\to^*_{\A} (s_1,\v_1, w_1)$ iff
$$\bigvee_{\eta\in\Phi}(
(\down{\v_0},\down{\v_1})\in\eta \land
(s_0,\up{\v_0}, w_0)\to^*_{\A,\eta} (s_1,\up{\v_1}, w_1)).$$
\end{lemma}
\begin{proof}
($\Rightarrow$) is immediate.

($\Leftarrow$) uses the following observation (from the
definition of
$\to^*_{\A,\eta}$ and Lemma \ref{strongbisim}):
for any pattern $\eta$,
$(\down{\v_0},\down{\v_1})\in\eta \land
(s_0,\up{\v_0}, w_0)\to^*_{\A,\eta} (s_1,\up{\v_1}, w_1)$
implies
$(\down{\v_0},\down{\v_1})\in\eta \land
(s_0,\up{\v_0}+\down{\v_0}, w_0)\to^*_{\A} (s_1,\up{\v_1}+\down{\v_1}, w_1)$.
\qed
\end{proof}

Once we give a characterization of
$\to^*_{\A,\eta}$, 
Lemma \ref{character} immediately gives a characterization for
$\to^*_{\A}$. Fortunately, the characterization of
$\to^*_{\A,\eta}$ is a decidable one, 
as shown in the next section.

\section{The Pattern Graph of a Timed Pushdown Automaton} \label{graph}

Let $\A=\langle S, \{x_1,\cdots,x_k\}, Inv, R, \Gamma, PD\rangle$
 be a PTA specified in the previous section.
The  {\em pattern graph} $G$ of $\A$ is a tuple
$$\langle S\times\Phi, \{y_0,\cdots,y_k\}, E, \Gamma\rangle$$
where
\begin{itemize}
\item $S$ is the states in $\A$,
\item $\Phi$ is the set of all patterns.
 A {\em node}
is an element in $S\times\Phi$,
\item {\em Discrete clocks}
$y_0,\cdots,y_k$
are the integral parts of the clocks
$x_0,\cdots,x_k$ in $\A$,
\item 
$E$ is a finite set of (directed) {\em edges} that connect
between nodes. 
An edge can be  a {\em progress} edge, a {\em stay}
edge,  or a 
{\em reset} edge.
A progress edge corresponds to progress transitions in $\A$
that cause one pattern change.
A stay edge  corresponds to progress transitions in $\A$
that cause no pattern change.
Since a progress transition
can cause no pattern change only from a  merge-pattern,
a stay edge connects a merge-pattern to itself.
A reset edge corresponds to a reset transition in $\A$.
Formally, 
 a progress edge $e_{s,\eta,\eta'}$
 that
connects node $(s,\eta)$ to
node $(s,\eta')$  is in the form of
$\langle (s,\eta), c,  (s,\eta')\rangle$
such that $c=Inv(s)$, $\eta'=Next(\eta)$ (thus $\eta\ne \eta'$).
A  stay  edge $e_{s,\eta,\eta}$, with $\eta$ being a merge-pattern,
 that
connects node $(s,\eta)$ to
itself  is in the form of
$\langle (s,\eta), c,  (s,\eta)\rangle$
such that $c=Inv(s)$. 
A  reset 
edge $e_{s,s',r, (a,\gamma)}$ that
connects node $(s,\eta)$ to  
node $(s',\eta')$  is in the form of
$$\langle (s,\eta), c, r, a, \gamma,  (s',\eta')\rangle$$
where 
$R(s,s')=(c,r)$ and $PD(s,s')=(a,\gamma)$. $E$ is the set of all 
progress edges, stay edges, and reset edges wrt $\A$. Obviously,
$E$ is finite.
\end{itemize}

A configuration of $G$ is a tuple $(s,\eta,\u, w)$
of state  $s\in S$, 
pattern $\eta\in\Phi$, 
discrete valuation $\u\in (\N^+)^{k+1}$ and 
stack word $w\in\Gamma^*$.
$(s,\eta, \u, w)\to^e (s',\eta', \u', w')$ denotes a {\em
one-step
transition}
 through edge $e$ of $G$ if the following conditions
are satisfied:
\begin{itemize}
\item if $e$ is a progress edge, then $e$ takes the form
$\langle (s,\eta), c, (s,\eta')\rangle$
and $s'=s$, $\u\in c^\eta$, $\u'\in c^{\eta'}$,
$next(\eta,\u)=(\eta',\u')$ and $w=w'$. 
Here $c^\eta$ and $c^{\eta'}$
are called the
{\em pre-} and the {\em 
post- (progress) tests} on edge $e$, respectively.
\item if $e$ is a stay  edge, then $e$ takes the form
$\langle (s,\eta), c, (s,\eta)\rangle$
and $s=s', \u\in c^\eta, \u=\u', \eta=\eta'$ and $w=w'$.
Here $c^\eta$ is called the
{\em pre-} and the {\em post- (stay
) tests} on edge $e$.
\item if $e$ is a reset  edge, then $e$ takes the form
$\langle (s,\eta), c, r,a,\gamma, (s',\eta')\rangle$
and $\u\in (c\land Inv(s))^\eta$, 
$\u'\in {Inv(s')}^{\eta'}$,
 $reset_r(\eta,\u)=(\eta',\u')$ and
$w=aw'', w'=\gamma w''$ for some $w''\in\Gamma^*$
(i.e., $w$ changes to $w'$ according to the stack operation). Here
$(c\land Inv(s))^\eta$ and ${Inv(s')}^{\eta'}$
are called the {\em pre-} and the {\em 
post- (reset) tests} on edge $e$, respectively.
\end{itemize}
We write $(s,\eta, \u, w)\to_G(s',\eta',\u', w')$
if $(s,\eta, \u, w)\to^e (s',\eta',\u', w')$ for some $e$.
The binary reachability
$\to_G^*$ of $G$
 is the transitive closure of $\to_G$.

The pattern graph $G$ simulates $\A$ in a way that
the integral parts of the dense
clocks are kept but
the fractional parts are abstracted as a pattern.
Edges in $G$ indicate how the pattern and the
discrete clocks change when a clock progress or
a clock reset occurs in $\A$.
However, a progress transition in $\A$ could cause
more than one pattern change. In this case, 
this big progress transition is treated as a 
sequence of small progress transitions
such that each causes one pattern change
(and therefore, each small progress transition
in $\A$
can be simulated by a progress edge in $G$). 
We first show that the binary reachability
$\to_G^*$ of $G$ is NPCA.
Observe that discrete clocks $y_0,\cdots, y_k$
are the integral values of dense clocks
$x_0,\cdots,x_k$.
Even though the dense clocks progress synchronously,
the discrete clocks may not be synchronous
(i.e., that
one discrete clock is incremented by 1
does not necessarily cause {\bf all} the other 
discrete clocks incremented by the same amount.).
The proof has two parts.
In the first part of the proof,
a technique is used to 
translate
$y_0,\cdots, y_k$ into
another array
 of discrete clocks that are synchronous.
In the second part of the proof,
$G$ can be treated as a discrete PTA
\cite{DIBKS00} by replacing $y_0,\cdots, y_k$
with the synchronous discrete clocks.
Therefore, Lemma \ref{Presburger} is obtained 
from 
the fact \cite{DIBKS00} that the binary reachability of
discrete PTA is NPCA.\footnote{For the purpose of this paper, we assume
in Lemma \ref{Presburger}
$\to_G^*$ is restricted in such a way that
$\eta$ is a regulated pattern whenever
$(s,\eta, \u, w)\to_G^*(s',\eta',\u', w')$. This
is because the auxiliary clock $x_0$ in $\A$ starts from 0.}

\begin{lemma}\label{Presburger}
For any PTA $\A$,
the binary reachability $\to_G^*$
 of the pattern graph $G$ of $\A$ 
is NPCA. In particular, if $\A$ is a timed automaton,
then the binary reachability $\to_G^*$ is Presburger.
\end{lemma}
\begin{proof}
We start with a technique that 
makes discrete clocks $y_0,\cdots,y_k$
(i.e., the integral parts of dense clocks)
synchronous on any path of $G$.

A {\em pattern ordering graph} $\P$ is a directed graph
on $\Phi$. For each (ordered) pair $(\eta,\eta')$ in $\Phi\times\Phi$,
$(\eta,\eta')$ is a progress edge, written $\eta\to_p\eta'$,
if $Next(\eta)=\eta'$.
In this case, we say the edge has label $p$
(stands for ``progress") and
 $\eta'$ is called the
{\em $p$-successor} of $\eta$.
$(\eta,\eta')$ is a reset edge with $r\subseteq K^+$,
written $\eta\to_r\eta'$,
if $Reset_r(\eta)=\eta'$.
In this case, we say the edge has label $r$ and
$\eta'$ is called the {\em $r$-successor} of $\eta$.
An edge can have multiple labels.

A {\em path} $\tau$
on $\P$ is a sequence of edges
$$\eta_0\to_{l_1}\cdots\to_{l_m}\eta_m$$
such that each $l_i$ is a label (either $p$ or some $r\subseteq K^+$).
 $j\in K^+$  {\em is reset on path $\tau$} if
$j\in l_i\subseteq K^+$ for some $1\le i\le m$.
Path $\tau$ is a {\em $p$-path} if
each edge on the path is a progress edge; i.e.,
label $l_i$ is $p$ for all $1\le i\le m$.
Path $\tau$ is a {\em regulated path}
if  $\eta_0$
is a regulated
pattern.
Path $\tau$ is  a {\em $p$-ring of $\eta_0$} if
$\tau$ is a $p$-path, and
$\eta_0,\cdots,\eta_m$ is the pattern ring of $\eta_0$.

Now we augment $\P$ with counters $\vec y$ ($=y_0,\cdots,y_k$)
taking values in $(\N^+)^{k+1}$.
Values of counters $\vec y$ change along a path in $\P$.
For each progress edge $\eta\to_p\eta'$, counters
$\vec y$ change to $\vec y'$ as follows:
$\vec y':=\vec y+\Delta_\eta$ (recall $\Delta_\eta$ is the increment vector for
$\eta$), consistent to the definition
that $next(\eta,\vec y)=(\eta', \vec y+\Delta_\eta)$.
For each reset edge $\eta\to_r\eta'$, counters
$\vec y$ change to $\vec y'$ as follows:
$\vec y':=\vec y\downarrow_r$, consistent to
the definition that $reset_r(\eta,\vec y)=(\eta', \vec y\downarrow_r)$.
For a $p$-path $\tau=\eta_0,\cdots,\eta_m$,
$\Delta_\tau=\Sigma_{0\le i\le m-1} \Delta_{\eta_i}$
is the net increment for counters $\vec y$ after walking through
the path. In particular,
$\Delta_\tau=\vec 1$ for each $p$-ring $\tau$.

A progress edge $\eta\to_p\eta'$
 is {\em add-$\vec 1$}
if $\eta'$ is a regulated pattern.
A path  is {\em short}
if it is a regulated path and,
it does not contain an add-$\vec 1$ edge
or it contains an add-$\vec 1$ edge but only at the end of the path.
A path  is {\em add-$\vec 1$} if
it is a short path containing an add-$\vec 1$ edge.
By definition, an add-$\vec 1$ path starts and ends with
regulated patterns and each pattern in between 
along the path
is not
a regulated pattern.
The following lemma is directly from the definitions
of $reset$ and $next$.

\begin{lemma}\label{add1}
For any path $\tau$, (1). if $\tau$ is a short path, then
for each $i\in K^+$ that is reset on $\tau$,
$y_i$ has value 0 at the end of $\tau$,
(2). if $\tau$ is an add-$\vec 1$ path, then
for each $i\in K^+$ that is not reset  on $\tau$,
$y_i$ has progressed by exactly 1  at the end of $\tau$.
\end{lemma}

When walking along a path in $\P$, a counter
in $\vec y$ is always nondecreasing except sometimes
it resets. However, counters $\vec y$ are not synchronous:
that one counter's advancing
 by 1 at some progress edge
 does not always cause {\bf all}
the other counters to advance by the same amount.

Now we are going
 to show that, on any regulated path,
 $\vec y$ can be simulated
by a set of synchronous  counters $\vec z=z_0,\cdots,z_k$.
The ideas are
 as follows.
Let $\tau$ be any regulated path of $\P$.
$\tau$ then can be concatenated by {\em segments}:
a number of
add-$\vec 1$ paths followed by a short path.
We introduce an increment vector $\Delta\in \{0,1\}^{k+1}$
to denote how much a counter in $\vec y$ progresses on a
segment.
Besides, we use $I\subseteq K^+$ to remember the indices $i\in K^+$
that are reset on each segment.
Assume counters $\vec y$ walk through $\tau$ and
change counter values from $\u$ to $\u'$.
Then, in the simulation, counters $\vec z$ starts from
$\u$ with $\Delta=\vec 0$ and $I=\emptyset$.
After walking through $\tau$ (while updating $\Delta$ and $I$
 along the path),
counters $\vec z$
have values satisfying
$\u'=(\vec z+\Delta)\downarrow_I$.
The simulation is defined by the following translation.
For each progress edge $\eta\to_p\eta'$,
the instruction $\vec y':=\vec y+\Delta_\eta$ is replaced by:
\begin{itemize}
\item if $\eta'$ is a regulated pattern (hence the edge is
an add-$\vec 1$ edge), i.e., the end of the current segment,
then
$\vec z':=(\vec z+\vec 1)\downarrow_I$ (synchronous progress followed by
resets);
$I':=\emptyset$;
$\Delta':=\bf 0$;
\item else,
$\vec z':=\vec z$;
$I':=I$;
$\Delta':=\Delta+\Delta_\eta$.
\end{itemize}
For each reset edge $\eta\to_r\eta'$,
the instruction $\vec y':=\vec y\downarrow_r$ is replaced by:
\begin{itemize}
\item $\vec z':=\vec z\downarrow_r$;
$\Delta':=\Delta\downarrow_r$;
$I':=I\cup r$.
\end{itemize}
Obviously $\vec z$ are synchronous. The correctness of the algorithm is
stated as follows.

\medskip

\noindent {\bf Claim.}
For any regulated path $\tau$, $\vec y=(\vec z+\Delta)\downarrow_I$
at the end of $\tau$.
\begin{proof}
Given a regulated path $\tau$.
Since $\tau$ can be split into a number of segments as mentioned
before, and by looking at the translation, at the end of
each add-$\vec 1$ path, $\Delta=\vec 0$ and $I=\emptyset$
(i.e., the initial values for $\Delta$ and $I$).
Therefore,
it suffices to show the claim for a segment, i.e., a short path $\tau$,
by induction on the length of $\tau$.
Notice that, from  the translation,
$I$ stands for the set of indices that has been reset on
the  short path; $\Delta$ stands for the increment 
that has been made
on the
 short path for counters $\vec y$. The relationship between
$I$ and $\Delta$ is established in Lemma \ref{add1}, which will be used
in the proof.

Case 1. The claim trivially holds for $\tau$ with length 1.

Case 2. Assume the claim holds for short paths with length $\le m$.
Now consider a short path
 with length $m+1$. This path can be written as a short path
$\tau$ followed by an edge $e$ of  $(\eta,\eta')$.
Note that, by the induction hypothesis,
 $\vec y=(\vec z+\Delta)\downarrow_I$ at $\eta$ (the end of
$\tau$). Now we are going to show
$\vec y'=(\vec z'+\Delta')\downarrow_{I'}$ where primed values are
for node $\eta'$.

Case 2.1. If  edge $e$ is a progress edge and
$\eta'$ is a regulated pattern, then, from the translation,
$\vec z'=(\vec z+\vec 1)\downarrow_I$,
$I'=\emptyset$,
$\Delta'=\bf 0$,
Therefore,

$\vec y'=\vec y+\Delta_\eta=$ (induction)

$(\vec z+\Delta)\downarrow_I+\Delta_\eta=$ (Lemma \ref{add1}(1))

$(\vec z+\Delta+\Delta_\eta)\downarrow_I=$ (Lemma \ref{add1}(2))

$(\vec z+\vec 1)\downarrow_I= \vec z'=$ (since $I'=\emptyset, \Delta'=\vec 0$)

$(\vec z'+\Delta')\downarrow_{I'}$.

Case 2.2. If the edge is a progress edge and
$\eta'$ is not a regulated pattern, then, from the translation,
$\vec z'=\vec z$,
$I'=I$,
and $\Delta'=\Delta+\Delta_\eta$.
Therefore,

$\vec y'=\vec y+\Delta_\eta=$ (induction)

$(\vec z+\Delta)\downarrow_I+\Delta_\eta=$ (Lemma \ref{add1}(1))

$(\vec z+\Delta+\Delta_\eta)\downarrow_I=$ (since $I'=I$,
and $\Delta'=\Delta+\Delta_\eta$)

$(\vec z'+\Delta')\downarrow_{I'}$.

Case 2.3. If the edge is a reset edge $\eta\to_r\eta'$,
then, from the translation,
$\vec z'=\vec z\downarrow_r$,
$\Delta'=\Delta\downarrow_r$, and
$I'=I\cup r$.
Therefore,

$\vec y'=\vec y\downarrow_r=$ (induction)

$(\vec z+\Delta)\downarrow_I\downarrow_r=$

$(\vec z'+\Delta')\downarrow_{I'}$.

Hence, the claim holds.
\qed
\end{proof}

Now we continue the proof of Lemma  \ref{Presburger}.
Let $G$
be the pattern graph of a timed automaton $\A$.
A path in $G$ witnessing
$$(s,\eta,\u, w)\to_G^* (s',\eta',\u', w')$$
(with $\eta$ being a regulated pattern)
between two configurations corresponds to
a regulated path (by properly adding stack operations)
 in the pattern ordering graph $\P$.
In above, we have demonstrated a technique
such that counters $\vec y=y_0,\cdots,y_k$ can be simulated
by synchronous counters $\vec z=z_0,\cdots,z_k$  using
an increment vector $\Delta\in\{0,1\}^{k+1}$ and a
reset set $I\subseteq K^+$.
The relationship between $\vec y$ and $\vec z$ is
$\vec y=(\vec z+\Delta)\downarrow_I$.
Tests
in $G$ 
(including
all the
pre- and post- (progress, stay and reset) tests)
 are in the form
of Boolean combinations of $y_i-y_j\# d$, $y_i\# d$ with $i,j\in K^+$
and $d\in \N^+$ (Section \ref{CCP}).
Since there are only a finite number of
choices for
$I$ and $\Delta$,
these tests can be accordingly translated to tests
on $z_0,\cdots,z_k$, using the relationship
$\vec y=(\vec z+\Delta)\downarrow_I$.
Observe that
the translated tests are
still
in the form of Boolean combinations of $z_i-z_j\# d$, $z_i\# d$ with
$i,j\in K^+$ and with probably larger or smaller $d$.
Since $\vec z$ are synchronous, $G$, with $\vec y$ simulated by $\vec z$,
is a discrete PTA
\cite{DIBKS00}. In that paper,
these synchronized discrete clocks $\vec z$
can be further translated into 
reversal-bounded counters. Hence,
the binary reachability of
a discrete PTA
is NPCA as shown in \cite{DIBKS00}.
 Therefore,
the lemma follows by
translating back from $\vec z$ to $\vec y$ using
$\vec y=(\vec z+\Delta)\downarrow_I$ at the initial and at the end
of the simulation (this requires only
a finite number of counter reversals).
Thus, $\to_G^*$ is NPCA.

In particular, when
$\A$ is a timed automaton, $G$, with $\vec y$ simulated by $\vec z$,
is a discrete timed automaton \cite{DIBKS00}.
Using the fact \cite{DIBKS00} that
the binary reachability of
a discrete timed automaton is Presburger,
$\to_G^*$ is also Presburger after the translation from
$\vec z$ back to $\vec y$.

\hfill\qed
\end{proof}

The following lemma states that $G$ faithfully simulates
$\A$ when the fractional parts of dense clocks 
are abstracted away by a pattern.
\begin{lemma}\label{Presburgersimulation}
Let $\A$ be a PTA with
pattern graph $G$.
For any states $s_0$ and $s_1$ in $S$,
any pattern
$\eta\in\Phi$, any stack words $w_0$ and $w_1$
in $\Gamma^*$,
 and any 
discrete valuation pairs
$(\u_0,\u_1)$ with $\u_0(0)=0$,
we have,
$$(s_0,\u_0, w_0)\to^*_{\A,\eta}(s_1,\u_1, w_1) \mbox{ ~iff~  }
 (s_0,init(\eta),\u_0, w_0)\to^*_G(s_1,\eta,\u_1, w_1).$$
\end{lemma}
\begin{proof}
Fix any
states $s_0,s_1\in S$,
any pattern
$\eta\in\Phi$, any
stack words $w_0$ and $w_1$
in $\Gamma^*$,
 and any
discrete valuation pairs
$(\u_0,\u_1)$ with $\u_0(0)=0$.

\noindent ($\Rightarrow$).
By the definition of $(s_0,\u_0, w_0)\to^*_{\A,\eta}(s_1,\u_1, w_1)$,
there exists an  initialized pair $(\v_0, \v_1)$
 such that
\begin{itemize}
\item $(\v_0, \v_1)$ has pattern $\eta$,
\item $\up{\v_0}=\u_0, \up{\v_1}=\u_1$,
\item $(s_0, \v_0, w_0)\to^*_{\A} (s_1,\v_1, w_1)$.
\end{itemize}
In order to show that
$(s_0,[(\v_0,\v_0)],\up{\v_0}, w_0)\to^*_G (s_1, 
[(\v_0,\v_1)],\up{\v_1}, w_1)$
(notice that $init(\eta)=[(\v_0,\v_0)]$),
it suffices to show that
each one-step transition in $\A$ can be simulated by
$\to^*_G$ properly: for any
valuations $\v,\v'$, any  states $s$ and $s'$,
and any stack words $w$ and $w'$,
if $(s,\v, w)\to_\A(s',\v', w')$ then
$(s, [(\v_0,\v)],\up{\v}, w)\to^*_G (s',[(\v_0,\v')],\up{\v'}, w')$.

Case 1. For any valuation $\v$ and state $s$, consider
a progress transition in $\A$, $(s,\v, w)$ $\to_\A(s,\v+\delta, w')$,
$\delta>0$,
such that (by definition) $w=w'$, and
$\forall 0\le\delta'\le \delta, v+\delta'\in Inv(s)$.
Let $\eta_0$  be the pattern of
$(\v_0,\v)$.
If $\v$ has no pattern change for $\delta$, then
$\eta_0$ must be a merge-pattern.
This progress transition in $\A$ can therefore be simply simulated
by the stay edge in $G$ at state $s$.
If, however, $\v$ has at least one pattern change for $\delta$, then
assume the $p$-ring of $\eta_0$ is
$\eta_0,\cdots,\eta_m=\eta_0$.
This progress transition in $\A$ can be simulated by
the following path consisting of progress edges
 in $G$:
looping along  the $p$-ring for $\up{\delta}$
times on state $s$ in $G$, followed by
a prefix of the $p$-ring ended with the pattern
$\eta_i$, for some $i$, of
$(\v_0,\v+\delta)$. From Lemma \ref{progresspatternchange}
and Lemma \ref{maxpatternchange},
it can be established $(s,\eta_0, \up{\v}, w)\to^*_G
(s,\eta_i,\up{\v+\delta}, w)$ through the path in $G$, noticing that
tests for $Inv(s)$ are consistent in $\A$ and $G$
(Lemma \ref{tests}), and the stack word does not change for progress
transitions in both $\A$ and $G$.

Case 2. For any valuation $\v$ and states $s$ and $s'$,
consider a reset  transition
$(s,\v, w)$ $\to_\A$ $(s', \v\downarrow_r, w')$  in $\A$
such that (by definition)
$w=aw'', w'=\gamma w''$ for some $w''$ with $PD(s,s')=(a,\gamma)$,
$R(s,s')=(c,r)$ and $\v\in Inv(s)\land c$, $\v\downarrow_r\in Inv(s')$.
Assume the pattern of $(\v_0,\v)$ is $\eta_0$ and
the pattern of $(\v_0,\v\downarrow_r)$ is $\eta_0'$.
This reset transition in $\A$ corresponds to the reset edge
in $G$:
$\langle (s,\eta_0), c, r, a, \gamma, (s',\eta_0')\rangle$.
 From Lemma \ref{resetpatternchange},
it can be established $(s,\eta_0, \up{\v}, w)\to^*_G
(s', \eta_0',\up{\v\downarrow_r}, w')$ through this edge,
noticing that
tests for $Inv(s)\land c$ and $Inv(s')$  are consistent in $\A$ and $G$
(Lemma \ref{tests}), and
the stack operations are the same in $\A$ and $G$.

\noindent ($\Leftarrow$).
Suppose
$(s_0,init(\eta),\u_0, w_0)\to^*_G(s_1,\eta,\u_1, w_1)$.
We would like to show
$$(s_0,\u_0, w_0)\to^*_{\A,\eta}(s_1,\u_1, w_1).$$
Pick any initial valuation $\v_0$ such that $(\v_0,\v_0)$ has
pattern $init(\eta)$ and $\up{\v_0}=\u_0$.
Suppose
$(s^0,\eta_0,\u^0, w^0)\to^{e_1}\cdots\to^{e_m} (s^m,\eta_m,\u^m, w^m)$
is a path (in $G$) witnessing
$(s_0,init(\eta),\u_0, w_0)\to^*_G(s_1,\eta,\u_1, w_1)$ through edges
$e_1,\cdots,e_m$
 such that $$(s^0,\eta_0,\u^0, w^0)=(s_0,init(\eta),\u_0, w_0)$$ and
$$(s^m,\eta_m,\u^m, w^m)=(s_1,\eta,\u_1, w_1).$$
A  path
in $\A$ 
$$(s^0,\v^0, w^0)\to^{t_1}\cdots\to^{t_m} (s^m,\v^m, w^m)$$
is constructed as follows, where
$\v^0=\v_0$ and each transition $t_i$ in $\A$
 corresponds to
each edge $e_i$ in $G$.
From $i=1$ to $m$, each $e_i$ belongs to one of the following 
three cases:

Case 1.
$e_i$ is a progress edge in $G$.
In this case, 
$next(\eta_{i-1},\u^{i-1})=(\eta_{i},\u^{i})$,
$w^{i}=w^{i-1}$,
and $s^{i-1}=s^{i}$.
We pick $t_i$ to be
a progress transition (at state $s^{i-1}$) in $\A$ from
$\v^{i-1}$ with an amount of
$\delta$ that causes exactly one pattern change
(Lemma \ref{maxpatternchange} and Lemma \ref{progresspatternchange}).
Take $\v^{i}=\v^{i-1}+\delta$.
Notice that
both the progress edge and the progress transition
do not change the stack content, i.e.,
$w^{i}=w^{i-1}$.

Case 2.
$e_i$ is a stay edge in $G$.
In this case, 
$\eta_{i-1}=\eta_{i}$ must be a merge-pattern with $w^{i}=w^{i-1}$
and
and $s^{i-1}=s^{i}$.
We pick $t_i$ to be
a progress transition (at state $s^{i-1}$) in $\A$ from
$\v^{i-1}$ with an amount of
$\delta$ that causes no pattern change
(Lemma \ref{maxpatternchange}).
Similarly to Case 1, $w^{i}=w^{i-1}$.

Case 3. $e_i$ is a
reset edge from state $s^{i-1}$ to state 
$s^{i}$ with clock resets $r$ in $G$,
then $t_i$ is the reset transition 
from state $s^{i-1}$ to state  
$s^{i}$ with
clock resets $r$ in $\A$. Notice that both $e_i$ and
$t_i$ have the same stack operation.
Take $\v^{i}=\v^{i-1}\downarrow r$
and $w^{i}$ is the result of the stack operation on $w^{i-1}$.

Notice that, for each $i=1\cdots m$,
\begin{itemize}
\item $(\v_0,\v^i)$ has pattern $\eta_i$,
\item $\up{\v^i}=\u^i$.
\end{itemize}
This can be shown using Lemma \ref{progresspatternchange} for Case 1,
Lemma \ref{maxpatternchange} for Case 2, and
Lemma \ref{resetpatternchange} for Case 3.
Therefore, this
 constructed path of $\A$ keeps the exactly
the same patterns and integral parts of
clocks as well as the stack word
 as in the path for $G$. 
From Lemma \ref{tests},
clock tests (and obviously the stack operations)
 are consistent between the path in $G$ and
the constructed path in $\A$.
Hence, 
$(s_0,\u_0, w_0)\to^*_{\A,\eta}(s_1,\u_1, w_1)$ since, by taking
$\v_1=\v^m$,
\begin{itemize}
\item $(\v_0, \v_1)$ has pattern $\eta$,
\item $\up{\v_0}=\u_0, \up{\v_1}=\u_1$,
\item $(s_0, \v_0, w_0)\to^*_{\A} (s_1,\v_1, w_1)$.
\end{itemize}
\qed
\end{proof}

Now, we conclude this section by claiming  that 
$\to^*_{\A,\eta}$ is NPCA by combining Lemma \ref{Presburger}
and Lemma \ref{Presburgersimulation}.
\begin{lemma}\label{integral}
For any PTA $\A$ and
any fixed pattern
$\eta\in\Phi$,
$\to^*_{\A,\eta}$ is NPCA.
In particular,
if $\A$ is a timed automaton,
then $\to^*_{\A,\eta}$ is Presburger.
\end{lemma}

\section{A Decidable Binary Reachability Characterization and 
Automatic Verification}\label{verif}

Recall that   PTA
$\A$ actually has clocks $x_1,\cdots, x_k$.
$x_0$ is the auxiliary clock.
The {\em binary reachability} $\leadsto^{*\B}_\A$
 of $\A$ is the set of
tuples
$$\langle s, v_1,\cdots, v_k, w,
s', v_1',\cdots,v_k', w'\rangle$$
such that
there exist $v_0=0, v_0'\in \D$
satisfying
$$(s, v_0, \cdots, v_k, w)\leadsto^*_\A
(s', v_0',\cdots,v_k', w').$$
The main theorem of this paper gives
a decidable characterization for the binary reachability
as follows.
\begin{theorem}\label{main}
The binary reachability $\leadsto^{*\B}_\A$
of a PTA $\A$ is $\FNPCM$-definable.
In particular, if $\A$ is a timed automaton,
then the binary reachability $\leadsto^{*\B}_\A$
can be expressed in the additive theory of reals (or rationals) and
integers.
\end{theorem}
\begin{proof}
From Lemma \ref{character},
$\leadsto^{*\B}_\A$ is definable by the following formula:
$$\exists u_0'\in \N^+
\exists v_0'\in \widehat\D(\bigvee_{\eta\in\Phi}
((0, v_1,\cdots, v_k),(v_0',\cdots,v_k'))\in\eta \land$$

$$ (s,(0,u_1, \cdots, u_k), w)\leadsto^*_{\A,\eta}
(s', (u_0',\cdots,u_k'), w'))
$$
on integer variables $s, u_1,\cdots, u_k, s', u_1', \cdots, u_k'$
(over $\N^+$), and dense variables
$v_1,\cdots,$ $v_k, v_1',\cdots, v_k'$ (over $\widehat\D=\D\cap [0,1)$), and
on word variables $w$ and $w'$.
This formula is equivalent to
$$\bigvee_{\eta\in\Phi} P_\eta^\D(v_1,\cdots, v_k,v_1',\cdots,v_k')
\land
Q_\eta^\N(s,u_1, \cdots, u_k, w,s',u_1',\cdots,u_k', w')
$$
where
$P_\eta^\D(v_1,\cdots, v_k,v_1',\cdots,v_k')$ stands for
$$\exists v_0'\in \widehat\D(((0, v_1,\cdots, v_k),(v_0',\cdots,v_k'))\in\eta )$$
and
$Q_\eta^\N(s,u_1, \cdots, 
u_k,w, s',u_1',\cdots,u_k', w')$ stands for
$$\exists u_0'((s,(0,u_1, \cdots, u_k), w)\leadsto^*_{\A,\eta}
(s', (u_0',\cdots,u_k'), w')).$$
From the definition of patterns, $P_\eta^\D$, after eliminating
the existential quantification, is a  dense linear
relation. On the other hand, $Q_\eta^\N$ (after eliminating
the existential quantification, from
Lemma \ref{integral} and  Lemma \ref{basicprop})
is NPCA. Therefore,
$\leadsto^{*\B}_\A$
is $\FNPCM$-definable.

In particular, if $\A$ is a timed automaton,
$\leadsto^{*\B}_\A$
is $\FNPCM$-definable by a formula
in the additive theory of reals (or rationals) and
integers.
Hence, $\leadsto^{*\B}_\A$ itself
can be expressed in the same theory.
\qed
\end{proof}
The importance of the above characterization for
$\leadsto^{*\B}_\A$ is that, from Lemma \ref{basicprop},
 the emptiness of 
$\FNPCM$-definable predicates
is decidable.
From Theorem \ref{main} and
Lemma \ref{basicprop} (3)(4), we have,
\begin{theorem}\label{verification}
The emptiness of $l~\cap\leadsto^{*\B}_\A$ 
with respect to a PTA $\A$ 
for any mixed linear relation $l$ is decidable.
\end{theorem}
The emptiness of $l~\cap\leadsto^{*\B}_\A$ is called a {\em mixed linear
property} of $\A$. Many interesting safety properties (or their 
negations) for
PTAs can be expressed as a mixed linear
property. For instance, consider
the following  property of a 
PTA $\A$ with three dense
clocks $x_1$, $x_2$ and $x_3$:

``for any two configurations $\alpha$ and $\beta$ with
$\alpha\leadsto^{*\B}_\A\beta$,
if the difference between $\beta_{x_3}$ (the
value of clock $x_3$ in $\beta$) and 
$\alpha_{x_1}+\alpha_{x_2}$ (the 
sum 
of clocks $x_1$ and $x_2$ in $\alpha$)
is greater than the difference between
$\#_a(\alpha_{\bf w})$ 
(the number of symbol $a$ appearing in the stack word in $\alpha$)
and 
$\#_b(\beta_{\bf w})$ (the number of symbol 
$b$ appearing in the stack word in $\beta$),
then $\#_a(\alpha_{\bf w})-2\#_b(\beta_{\bf w})$ is greater than 5."

\noindent The negation of this property can be 
expressed as the emptiness of 
$$(s, x_1,x_2, x_3, w)\leadsto^{*\B}_\A(s', x_1', x_2', x_3', w')\land
l$$
where $l$ is the negation
of a mixed linear relation (hence $l$ itself is also a
mixed linear relation):
$$x_3'-(x_1+x_2)>\#_a(w)-\#_b(w')
~\to~ \#_a(w)-2\#_b(w')>5.$$
Thus, from Theorem \ref{verification},
this property can be automatically verified.
We need to point out that
\begin{itemize}
\item $x_3'-(x_1+x_2)>\#_a(w)-\#_b(w')$ is a linear relation on both
dense variables and discrete variables. Thus, 
this property can not be verified by using  the decidable characterization
for discrete PTAs \cite{DIBKS00}, where only 
integer-valued clocks are considered. 
\item Even without clocks, $\#_a(w)-2\#_b(w')>5$ expresses a non-regular
set of stack word pairs. Therefore, this property can not be verified
by the model-checking procedures for pushdown systems \cite{BEM97,FWW97}.
\item Even without the pushdown stack,
$x_3'-(x_1+x_2)>0$ (by taking $\#_a(w)-\#_b(w')$ as
a constant such as 0)
is not a clock region, therefore, the classical 
region-based techniques can not
verify this property. This is also pointed out in \cite{CJ99}.
\item With both dense clocks and the pushdown stack,
this property can not be verified by using the region-based techniques 
for Timed Pushdown Systems \cite{BER95}.
\end{itemize}
When $\A$ is a timed automaton, by Theorem \ref{main},
the binary reachability $\leadsto^{*\B}_\A$ can be expressed
in the additive theory of reals (or rationals) and integers.
Notice that our characterization is essentially
equivalent to the one given by
Comon and Jurski \cite{CJ99} in which
$\leadsto^{*\B}_\A$ can be expressed
 in the additive theory of reals augmented with
a predicate telling whether a term is an integer.
Because the additive theory of reals and integers
is  decidable \cite{B62,BRW98}, we have,

\begin{theorem}\label{comon}
The truth value for any closed formula expressible in
the (first-order) additive theory of reals (or rationals)
augmented with a predicate $\leadsto^{*\B}_\A$ for a
timed automaton $\A$ is decidable. (also shown in \cite{CJ99})
\end{theorem}
For instance, consider the following property
for a timed automaton $\A$ with two real clocks:

``there are states $s$ and $s'$ such that,
for any $x_1, x_2, x_2'$, there exists $x_1'$ such that
if $(s, x_1, x_2)$ can reach $(s', x_1', x_2')$ in $\A$,
then $x_1-x_2>x_1'-x_2'$."

\noindent It can be expressed as
$$\exists s, s' \forall x_1, x_2, x_2' \exists
x_1' ((s, x_1, x_2)\leadsto^{*\B}_\A(s', x_1', x_2')
\to x_1-x_2>x_1'-x_2'),$$
and thus can be verified
according to Theorem \ref{comon}.

\section{Conclusions, Discussions and Future Work}\label{concl}


In this paper, we consider
PTAs that are timed automata
augmented with a pushdown stack.
A configuration of a  PTA
includes a control state, finitely many dense clock values
and a stack word.
By introducing the
 concept of a clock pattern
and using an automata-theoretic approach,
we give a decidable
characterization of
the binary reachability
 of a
PTA.
Since a timed automaton  can be treated as
a PTA
 without the pushdown stack,
we can show that the binary reachability of a timed automaton
is definable in the additive theory of reals and integers.
The results can be used to verify a class
 of safety properties containing linear relations
over both dense variables and unbounded discrete variables.

A PTA studied here
can be regarded as the timed version of
a pushdown machine. 
Carefully looking at the proofs of the decidable 
binary reachability characterization,
we  find out that the underlying
untimed machine (e.g., the pushdown machine)
is not essential.
We can replace it
with many other kinds of machines and
the resulting timed system still has
a decidable binary reachability characterization.
We will summarize some of these machines
in this section.

Consider a class of machines {\bf X}.
We use {\bf XCM} to denote machines
in {\bf X}
augmented with reversal-bounded counters.
We are looking at the binary reachability characterization
of the timed version of machines in {\bf X}.
The characterization is established in the previous sections
when {\bf X} represents pushdown machines.
In the proofs,
a dense clock is separated into a fractional part and
an integral part. The fractional parts of dense clocks
are abstracted as a pattern and the
integral parts are translated into synchronous discrete
clocks, which are further translated into
reversal-bounded counters \cite{DIBKS00}.
The result of the translation is 
the underlying untimed machine in {\bf X}
 augmented with these
reversal-bounded counters, i.e.,
a machine in {\bf XCM}.
Suppose a class of automata {\bf Y}
accept the binary reachability  of
machines in {\bf XCM}.
In the case of {\bf X} being pushdown machines,
{\bf XCM} represents NPCMs and
{\bf Y} can be chosen as NPCAs
(it is known that the binary reachability of
NPCMs can be accepted by NPCAs \cite{DIBKS00}.).
The fact that this {\bf Y} (i.e., NPCA)
satisfies Lemma \ref{basicprop}
is the only condition we need
in order to obtain the decidable reachability characterization
in Theorem \ref{main}.
Definitions like NPCA predicates and $\FNPCM$-definability
can be accordingly
modified into 
{\bf Y} predicates and ({\bf D+Y})-definability
 once {\bf Y} is clear.
The above discussions 
give the following result.

\begin{theorem}\label{generaldense}
Let {\bf Y} be a class of automata,
{\bf X} be a class of machines
and {\bf XCM} be the class of machines in {\bf X} 
augmented with reversal-bounded counters.
If, for each machine in {\bf XCM}, an automaton in
{\bf Y} can be constructed that accepts the binary
reachability of
the machine, and
Lemma \ref{basicprop} holds (replacing NPCA with {\bf Y}),
then
the binary reachability of the timed version of
{\bf X} is ({\bf D+Y})-definable.
\end{theorem}

Notice that Lemma \ref{basicprop} (4) requires that
the emptiness problem for {\bf Y} in Theorem \ref{generaldense}
be decidable. 
Theorem \ref{verification}
can be immediately followed from
Theorem \ref{generaldense} for
the timed version of {\bf X}.

According to Theorem \ref{generaldense},
the timed
version of the following machines {\bf X}
has a decidable ({\bf D+Y})-definable
characterization for binary reachability
by properly choosing {\bf Y}:
\begin{itemize}
\item NPCM. Here {\bf Y}=NPCA;
\item NCM with an unrestricted counter.
Notice that the counter is a special case of a pushdown stack
(when the stack alphabet is unary).  
Here, {\bf Y}=NPCA;
\item Finite-crossing NCM \cite{I78}
(i.e., NCM augmented with a finite-crossing read-only worktape.
The head on the worktape is two-way, but
for each cell of the tape, the head crosses only a bounded number of times.).
Here, {\bf Y} is finite-crossing NCAs
\cite{I78} that are NCM augmented with a finite-crossing input tape.
\item Reversal-bounded multipushdown machines \cite{D00}
that are multipushdown machines \cite{Italy}
augmented with reversal-bounded counters.
Here, {\bf Y} is reversal-bounded multipushdown automata \cite{D00}.
\end{itemize}

Let {\bf X} be a class of machines.
The pattern technique tells us that,
for a
decidable
binary reachability characterization of the timed version of
{\bf X}, the density of clocks (and even clocks themselves)
 is not the key issue.
This is because, using the technique, these dense clocks 
can be reduced to reversal-bounded integer counters.
The key issue is whether {\bf X}  and 
its reversal-bounded
version {\bf XCM}
have a decidable binary reachability characterization
(i.e., the binary reachability can be accepted
by a class {\bf Y} of automata with a
decidable emptiness problem).
In particular,
when the binary reachability of {\bf X} is 
effectively semilinear (and hence the binary reachability
is decidable), in most cases,
the binary reachability of {\bf XCM} is also
effectively semilinear.
Such {\bf X} includes all the machines mentioned above.
In this case,
once we can show the untimed machines
in {\bf X} have a decidable binary reachability characterization,
we are getting really close to the 
decidable characterization for their timed version.
But, we do have exceptions.
For instance, consider {\bf X} to be
a finite state machine with a two-way read only worktape.
{\bf X} has a decidable binary reachability characterization
(witnessed by  two-way multitape finite automata).
However, 
augmenting {\bf X} with reversal-bounded counters
makes the binary reachability undecidable.
The pitfall here is that a two-way tape makes
reversal-bounded counters too powerful.
In fact, the emptiness problem is undecidable
for two-way automata augmented with reversal-bounded
counters. In the case when there is only one reversal-bounded
counter, the emptiness problem is decidable if
the machines are deterministic.
The nondeterministic case is still open \cite{IJTW95}.

In practice, augmenting timed automata with other unbounded
data structures allows us to study more complex
real-time applications. For instance, the decidable characterization
of PTAs makes it possible to implement a tool
verifying recursive real-time programs containing
finite-state variables against safety properties
containing linear constraints over dense clocks and
stack word counts. This tool will be
a good complement to available tools
for recursive finite state programs
(for {\em regular} safety properties, e.g., 
termination) \cite{ES01,BR00}.
On the other hand,
for the existing tools
analyzing real-time systems
(such as UPPAAL \cite{LPY97} and its extensions \cite{LB01},
TREX \cite{LB01}, HyTECH \cite{HH95}, Kronos \cite{BDM98}),
the traditional
region-based technique used in the tools
 may be enhanced with the pattern technique. Doing this makes it
possible for the tools to verify complex timing requirements
that may not be in the form of clock regions.
The results in this paper can also be used to implement a
model-checker for a subset of
the real-time specification language ASTRAL
\cite{CGK97}. The subset includes history-independent
ASTRAL
specifications containing both
dense clocks and unbounded discrete control variables.

As mentioned in this section, 
the timed version of NPCM (i.e.,
PTAs further augmented with reversal-bounded counters)
also has a decidable characterization.
This timed model has many important applications.
For instance, a 
real-time recursive program (containing
unbounded integer variables)
can be automatically debugged using the reversal-bounded
approximation (i.e., assign a reversal-bound to the variables).
Additionally, a free counter (i.e., an unrestricted counter)
is a special case for a pushdown stack (when the stack alphabet is
unary). Therefore, this model can also be used to
specify real-time systems containing 
a free counter and many reversal-bounded counters.
It seems that ``reversal-bounded counters"
appear unnatural and therefore 
their applications in practice are remote.
However, a non-decreasing counter is also a reversal-bounded counter
(with zero reversal-bound).
This kind of counters have a lot of applications. For instance,
a non-decreasing counter can be used to
count
digital time elapse, the number of external events,
the number of a particular branch taken by a nondeterministic program
(this is important, when fairness is taken into account), etc.
For instance, consider a timed automaton with input symbols
(i.e., a transition is triggered by an external event as
well as the enabling condition).
We use $\#_a$ to denote the number of event $a$ occurred so far.
The enabling condition of a transition, besides clock constraints,
may also include comparisons of the counts $\#_a$ against an integer constant
and comparisons of one specific linear term $T$ (on all $\#_a$)
against an integer constant.
For instance, a transition may look like this (in pseudo-code):
      
{\tt $s$: if event($a$) and $x_2-x_1>10$ and
$\#_b>21$ and $2\#_c-3\#_b<5$, then progress(); goto $s'$}

\noindent where $x_1$ and $x_2$ are dense clocks. Notice that
comparisons of the linear term $2\#_c-3\#_b$ against
an integer constant
may show up in other transitions.
But this term is unique in the automaton:
a comparison like
$4\#_a-3\#_b>8$ that involves
a different term $4\#_a-3\#_b$
 can not be used in 
the enabling conditions of the automaton.
This timed automaton is a standard timed automaton
augmented with reversal-bounded counters $\#_a$
(which are non-decreasing) and
a free counter (representing the linear term $2\#_c-3\#_b$).
Hence, the following property can be automatically verified:

``It is always true that whenever $x_1-7\#_b+3x_2>2\#_a$
holds, $x_1$ must be greater 
$\#_c-\#_a$."

A future research issue is to
investigate whether the decidable results
\cite{DPK01} for Presburger liveness
of discrete timed automata can be 
extended to
timed automata (with dense clocks) using the technique in this paper.
We are also going to look at 
the possibility of extending the approximation
approaches for parameterized discrete timed automata \cite{DIK01}
to the dense clocks. This is particularly interesting,
since the reachability set presented in \cite{DIK01} is
not necessarily semilinear.
Another 
 issue is
on the complexity analysis
 of the decision procedure
presented in this paper.
However, the complexity for the emptiness problem of NPCAs
is still unknown, though it is believed that
it can be derived along Gurari and Ibarra \cite{GI81}.

The author would like to thank 
H. Comon and O. H. Ibarra
for discussions on the topic of dense timed
pushdown automata during CAV'00 in Chicago,
B. Boigelot, P. San Pietro and
J. Su for recent discussions on \cite{BRW98},
J. Nelson, F. Sheldon and
G. Xie for reading an earlier draft of this paper.
Thanks also go to T. Bultan, H. Comon,
J. Esparza and K. Larsen for comments on the short
version of this paper presented
 in
CAV'01 in Paris.


\begin{thebibliography}{20}

\bibitem{A99} R. Alur, {\sl ``Timed automata"},
 {\em CAV'99},
LNCS 1633, pp. 8-22



\bibitem{ACD93} R. Alur, C. Courcoibetis, and D. Dill,  {\sl
``Model-checking in dense real time,"}
{\em Information and Computation}, {\bf 104} (1993) 2-34


\bibitem{AD94}       R. Alur and D. Dill,  {\sl ``A theory of
timed automata,"}  {\em Theoretical Computer  Science},
{\bf 126} (1994) 183-236



\bibitem{AFH96}
R. Alur, T. Feder, and T. A. Henzinger,  {\sl ``The benefits of relaxing punctuality,"}  {\em J. ACM},
{\bf 43} (1996) 116-146


\bibitem {AH93} R. Alur, T. A. Henzinger,  {\sl ``Real-time 
logics: complexity and expressiveness,"}
{\em Information and Computation},
{\bf 104} (1993) 35-77



\bibitem {AH94} R. Alur, T. A. Henzinger,  {\sl ``A really
temporal logic,"} {\em J. ACM},
{\bf 41} (1994) 181-204


\bibitem {BR00}
T. Ball and S. K. Rajamani,
 {\sl
``Bebop: a symbolic model-checker for
Boolean programs,"}
{\em Spin Workshop'00}, LNCS 1885,
pp. 113-130.




\bibitem{B62}
J. R. Buchi,
{\sl ``On a decision method in restricted second
order arithmetic,"} 
{\em 
Proceedings of the International Congress on
Logic, Method, and Philosophy of Sciences,}
Stanford University Press, pp. 1-12, 1962



\bibitem{BEM97}
A. Bouajjani, J. Esparza, and O. Maler,  {\sl
``Reachability Analysis of Pushdown Automata: Application to Model-Checking,"},
{\em CONCUR'97}, LNCS 1243,  pp. 135-150


\bibitem{BER95}
A. Bouajjani, R. Echahed, and R. Robbana,  {\sl
``On the automatic verification of systems
                  with continuous variables
                  and unbounded discrete data structures,"}
{\em Hybrid System II},
LNCS 999, 1995, pp. 64-85

\bibitem{BDM98}
M. Bozga, C. Daws, O. Maler, A. Olivero, S. Tripakis, and
 S. Yovine,  {\sl
``Kronos: A model-checking tool
for real-time systems,"}
{\em CAV'98},
LNCS 1427, pp. 546-550




\bibitem {BRW98}
B. Boigelot, S. Rassart and P. Wolper,
 {\sl
``On the expressiveness of real and integer arithmetic automata,"}
{\em ICALP'98}, LNCS 1443, pp. 152-163




\bibitem{Italy} A. Cherubini, L. Breveglieri, C. Citrini,
and S. Crespi Reghizzi.
``Multi-push-down languages and
grammars,"
{\sl
International Journal of Foundations of Computer
     Science,
}
7(3): 253-291, 1996



\bibitem{CGK97}        A.\ Coen-Porisini, C.\ Ghezzi and R.\ Kemmerer,  {\sl
``Specification  of real-time systems using ASTRAL,"}
{\em IEEE Transactions on\  Software Engineering},
{\bf 23} (1997) 572-598

\bibitem{CJ98}
H. Comon and Y. Jurski,  {\sl
``Multiple counters automata, safety analysis and Presburger arithmetic,"}
{\em CAV'98}, LNCS 1427, pp. 268-279.


\bibitem{CJ99}
H. Comon and Y. Jurski,  {\sl
``Timed automata and the theory of real numbers,"}
{\em CONCUR'99}, LNCS 1664, pp. 242-257


\bibitem{D00}
Z. Dang, {\sl
``Debugging and verification of infinite state real-time systems,"}
PhD Dissertation, University of California, Santa Barbara, August 2000


\bibitem{D01}
Z. Dang, {\sl
``Binary reachability analysis of pushdown timed automata
with dense clocks,"}
 {\em CAV'01}, LNCS 2102, pp. 506-517






\bibitem{DIBKS00}
Z. Dang, O. H. Ibarra,  T. Bultan,
R. A. Kemmerer, and J. Su,  {\sl
``Binary reachability analysis of discrete
pushdown timed automata,"}
 {\em CAV'00}, LNCS 1855, pp. 69-84


\bibitem{DIK01}
Z. Dang, O. H. Ibarra and R. A. Kemmerer,  {\sl
``Decidable approximations on generalized and parameterized 
discrete timed automata,"}
 {\em COCOON'01}, LNCS 2108, pp. 529-539

 

\bibitem{DPK01}
Z. Dang, P. San Pietro and R. A. Kemmerer,  {\sl
``On Presburger liveness of discrete timed automata,"}
 {\em STACS'01}, LNCS 2010, pp. 132-143


\bibitem{ES01}
J. Esparza and S. Schwoon, {\sl
``A BDD-based model-checker for recursive programs,"}
{\em CAV'01},
LNCS 2102, pp. 324-336


\bibitem{FWW97}                    A. Finkel,
B. Willems and P. Wolper, {\sl
``A direct symbolic approach to model checking pushdown systems,"}
{\em INFINITY'97}.

\bibitem{GI81}       E. Gurari and O. Ibarra,
{\sl
``The complexity of decision problems for
       finite-turn multicounter machines,"}
{\em J. Computer and
       System Sciences}, {\bf 22} (1981) 220-229


\bibitem{HMP92}
T. A. Henzinger, Z. Manna, and A. Pnueli,
 {\sl ``What good are digital clocks?,"}
 {\em ICALP'92}, LNCS 623, pp. 545-558


\bibitem{HH95}
T. A. Henzinger and Pei-Hsin Ho,
 {\sl ``HyTech: the Cornell hybrid technology tool,"}
 {\em Hybrid Systems II},
LNCS 999, pp. 265-294



\bibitem{HNSY94}                   T. A. Henzinger, X. Nicollin, J. Sifakis,
 and S. Yovine. {\sl ``Symbolic model checking for real-time systems,"}
{\em Information and Computation}, {\bf 111} (1994) 193-244


\bibitem{I78} O. H. Ibarra,  {\sl
``Reversal-bounded multicounter machines and their decision problems,"}
{\em J. ACM}, {\bf 25} (1978) 116-133


\bibitem{IJTW95}
  O. H. Ibarra, T. Jiang, N. Tran and H. Wang,
  ``New decidability results concerning two-way counter machines,''
  {\em SIAM J. Comput.,} {\bf 24} (1995) 123-137








\bibitem{LPY97}
K. G. Larsen, P. Pattersson, and W. Yi,
 {\sl ``UPPAAL in a nutshell,"}
{\em International Journal on Software Tools for Technology Transfer},
{\bf 1} (1997): 134-152


\bibitem{LB01}
K. G. Larsen,
G. Behrmann, Ed Brinksma, A. Fehnker, T. Hune, 
P. Pettersson, and J. Romijn,
 {\sl ``As cheap as possible: efficient
     cost-optimal reachability for priced timed automata,"}
{\em CAV'01}, 
LNCS 2102, pp. 493-505


\bibitem{LLW95}
F. Laroussinie, K. G. Larsen, and
C. Weise,
 {\sl ``From timed automata to logic - and back,"}
 {\em MFCS'95}, LNCS 969, pp. 529-539


\bibitem{RS97}
J.  Raskin and P.  Schobben, {\sl ``State
clock logic: a decidable real-time logic,"}
{\em HART'97}, LNCS 1201, pp. 33-47
 


\bibitem{W94}
T. Wilke,
{\sl ``Specifying timed state sequences in powerful decidable logics
and timed automata,"}
LNCS 863, pp. 694-715, 1994




\bibitem{Y98}
S. Yovine,
 {\sl ``Model checking timed automata,"}
 {\em Embedded Systems'98}, LNCS 1494, pp. 114-152



\end{thebibliography}
\end{document}